\documentclass[aps,prd,twocolumn,superscriptaddress,amsmath,amssymb,floatfix,nofootinbib,longbibliography]{revtex4-2}
\usepackage{bm,graphicx,mathtools,stmaryrd,mathrsfs}
\usepackage[dvipsnames]{xcolor}
\usepackage{hyperref}
\hypersetup{colorlinks=true,linkcolor=NavyBlue,citecolor=ForestGreen,urlcolor=NavyBlue}

\newcommand{\tg}{\tilde g}
\newcommand{\tG}{\tilde{\mathbf G}}
\newcommand{\tM}{\tilde{\mathbf M}}
\newcommand{\tR}{\tilde{\bm{\mathcal R}}}
\newcommand{\tGc}{\tilde{\bm{\mathcal G}}}
\newcommand{\tTh}{\tilde{\bm\Theta}}
\newcommand{\tI}{\tilde{\mathbf I}}
\newcommand{\tN}{\tilde{\mathbf N}}
\newcommand{\tX}{\tilde{\mathbf X}}
\newcommand{\bG}{\mathbf G}
\newcommand{\bM}{\mathbf M}
\newcommand{\TrF}{\mathrm{Tr}_F}
\newcommand{\TrM}{\mathrm{Tr}_M}
\newcommand{\Tr}{\mathrm{Tr}}
\newcommand{\sg}{\sqrt{-g}}
\newcommand{\dr}{d^dx}
\newcommand{\lP}{\ell_P}
\newcommand{\cN}{\mathcal N}
\newcommand{\av}[1]{\langle #1\rangle}
\newcommand{\RG}{\mathcal R_{\mathcal G}}
\newcommand{\MG}{\mathcal M_{\mathcal G}}
\newcommand{\LG}{\Lambda_{\mathcal G}}
\newcommand{\ket}[1]{|#1\rangle}
\newcommand{\bra}[1]{\langle #1|}
\newcommand{\beff}{\beta_{\rm eff}}
\newcommand{\rev}[1]{#1}
\newcommand{\revm}[1]{#1}

\newcounter{hyp}
\renewcommand{\thehyp}{A\arabic{hyp}}
\newenvironment{hypothesis}[2]{\par\smallskip\noindent\refstepcounter{hyp}\label{#2}(\thehyp)~\textit{#1.}~\ignorespaces}{\par\smallskip}

\begin{document}

\title{Gravity from generalized entropy}

\author{Hussain Gohar}
\email{hussain.gohar@usz.edu.pl}
\affiliation{Institute of Physics, University of Szczecin,
             Wielkopolska 15, 70-451 Szczecin, Poland}

\date{\today}

\begin{abstract}
Gravity from entropy (GfE) identifies \rev{its Lagrangian} with the \rev{geometric quantum relative entropy (GQRE)} between the spacetime metric and a metric induced by matter and curvature\rev{, treating the metric itself as a quantum operator}. We extend this construction to the Tsallis, R\'enyi, and Sharma--Mittal families\rev{, which we call gravity from generalized entropy (GfGE),} and determine their common Einstein limit, weak-field spectra, auxiliary-field formulation, and horizon entropy. The entropy choice controls the leading higher-curvature terms and additional modes. \rev{We confirm that the linearized GfE spectrum} contains a healthy scalar and a Planck-scale tachyonic spin-2 ghost. We identify the conditions that remove this ghost: curvature must enter through its scalar trace, or the entropy generator must be flat to quadratic order at the identity. The latter gives exact general relativity for Tsallis at $q=0$ and ghost-free $f(R,\mathcal L_m)$ theories for R\'enyi and Sharma--Mittal. On shell, the auxiliary G-field is proportional to the tensor in Wald's entropy formula. For analytic entropy functions, horizons with a single curvature scale have an area term followed by constant and inverse-area corrections. We also extend the volume-constraint cosmological constant to the full family; its leading entropy dependence is set by the scalar mass. A finite set of entropy-function derivatives controls the leading weak-curvature departures from general relativity.
\end{abstract}


\maketitle

\section{Introduction}\label{sec:intro}
The relation between gravity, thermodynamics, and information has been explored since black holes were found to carry entropy and radiate~\cite{Bekenstein1973,Bekenstein1974,Hawking1975}. The Einstein equations can be derived from the Clausius relation on local horizons~\cite{Jacobson1995,Jacobson2016,Padmanabhan2010}, while holography relates entanglement entropy to geometry~\cite{tHooft,Susskind1995,RT2006,Swingle2012,FLM2013}. A complementary starting point is operator-algebraic quantum field theory, where Araki's relative entropy~\cite{Araki1976,OhyaPetz} is defined for local algebras without reference to a horizon and features in modern accounts of entanglement in quantum field theory and gravity~\cite{Witten2018,Sorce2023,Witten2022}. \rev{Recently, Dorau and Much showed that the semiclassical Einstein equations follow when the Araki--Uhlmann relative entropy between the vacuum and coherent excitations of a quantum field, which equals the energy flux through a local Rindler horizon, is identified with a quarter of the variation of the horizon area~\cite{DorauMuch2026}. There, the relative entropy compares states of a quantum field across a horizon; in GfE, it compares the metric with an induced metric, both treated as operators, and serves as a local action.}

Gravity from entropy (GfE), proposed by Bianconi~\cite{GfE}, identifies \rev{the Lagrangian with the quantum relative entropy between the spacetime metric and an induced metric, now called the geometric quantum relative entropy (GQRE)~\cite{GfEthermo,BianconiLambda2026}. Unlike Araki's relative entropy, which compares quantum states, the GQRE treats the metric itself as a quantum operator. This is the central physical insight of GfE, on which the present work builds: it brings the geometry of general relativity and quantum mechanics together already at the level of first quantization~\cite{GfE,GfEthermo}. The induced metric} is built from a complex Dirac--K\"ahler boson~\cite{Kahler1962,BecherJoos1982,Kruglov2002} (a 0-, 1-, and 2-form) and the curvature. In the weak-coupling limit, the action reduces to Einstein gravity without a cosmological constant. An auxiliary \emph{G-field} recasts it as a dressed Einstein--Hilbert action with an \rev{emergent, dynamical cosmological constant $\LG$~\cite{GfE}. GfE extends to the continuum an earlier construction on discrete higher-order networks~\cite{Bianconi2024}. Subsequent work by Bianconi derived the area law from the quantum relative entropy of the Schwarzschild black hole~\cite{BianconiBH2025}, developed the thermodynamics of the theory, identifying the energy density of the G-field with a dynamical dark-energy term~\cite{GfEthermo}, and introduced a local volume constraint that adds a constant cosmological term~\cite{BianconiLambda2026}.} \rev{Inflation~\cite{ThattarampillyZheng2025} and spherically symmetric black holes~\cite{Thattarampilly2026} have also been studied, and the vacuum spectrum and stability were analyzed by Pereira~\cite{Pereira2026a,Pereira2026b}.}

The logarithm is not the only possible entropy functional. Replacing it changes the higher-curvature terms, propagating modes, and horizon entropy, raising the question of which predictions are structural and which depend on the \rev{logarithmic (Umegaki)} choice. We answer this by constructing \rev{the actions of gravity from generalized entropy (GfGE)} from Tsallis~\cite{Tsallis1988,TsallisBook,Abe2003,Furuichi2004}, R\'enyi~\cite{Renyi1961,Petz1986,MullerLennert2013,WildeWinterYang2014}, and Sharma--Mittal~\cite{SharmaMittal1975} relative entropies, each of which recovers the Umegaki form at particular index values. These entropies have been used to modify horizon laws~\cite{TsallisCirto2013,CzinnerIguchi2016,Barrow2020,Saridakis2020} and in cosmological models~\cite{Tavayef2018,NOF2022}. \rev{A generalized mass-to-horizon relation, combined with the Clausius relation and the Hawking temperature, yields a broad family of such entropies consistently with thermodynamics~\cite{GoharMH2026}, and these entropies can be mapped onto scalar-tensor gravity with a scale-dependent effective gravitational coupling~\cite{GoharST2026}.} Here they define local actions, letting us test which proposed horizon modifications follow from the underlying dynamics. \rev{The generalization has a conceptual cost. In GfE the Lagrangian is a quantum relative entropy that also takes the form $\ln W$ of a Boltzmann--Gibbs entropy, so that the action can be read as the entropy of the geometric degrees of freedom of spacetime~\cite{GfEthermo}. This second reading rests on the logarithm and is lost for the generalized entropies (Sec.~\ref{sec:logdet}).} This paper develops the theories' foundations; a companion paper treats black holes, cosmology, and observational constraints~\cite{PaperII}.

Our results separate properties fixed by the construction from those controlled by the entropy. The common normalization gives every member the same Einstein--Hilbert term, Newton constant, and leading-order matter coupling. The non-negative divergence starts at quadratic order and supplies the entropy-dependent higher-curvature corrections; nonlinear deformations can, however, generate nonminimal matter couplings.

Around Minkowski space, the spectrum contains the graviton, a scalar, and generically a massive spin-2 mode. The latter is always a ghost when present and is tachyonic in GfE, with a Planck-scale mass\rev{, as first shown by Pereira~\cite{Pereira2026a}}. This establishes a linearized instability. \rev{Reference~\cite{GfE} asked whether the theory is affected by the Ostrogradsky instability and noted that its G-field equations are of second order; the spectrum shows that this does not remove the massive spin-2 pole, while the nonlinear question remains open.} For a common generating function, the massive spin-2 mode is absent only when curvature enters through its scalar trace or the generator is flat to quadratic order at the identity. In the Tsallis family, the latter occurs at $q=0$: Tsallis then gives general relativity, whereas R\'enyi and Sharma--Mittal give ghost-free $f(R,\mathcal L_m)$ theories.

On shell, the G-field is proportional to $\partial\mathcal L/\partial R_{\mu\nu\rho\sigma}$, the tensor entering both the field equations and Wald entropy. Horizon entropy is thus the area law weighted by the G-field components normal to the horizon. For a single horizon curvature scale, analyticity gives an area term followed by a constant and inverse-area corrections, rather than the power laws often postulated in nonextensive cosmology. We also extend the volume-constraint cosmological constant to the full family; at first order, entropy dependence in its trace equation is captured by the scalar mass.

At low curvature, only a few derivatives of the entropy functions at the identity control the leading corrections. At fixed $q$, the three families share the terms set by the generator and differ through the nonlinear coupling between form sectors.

The construction is distinct from spectral actions, whose traces are functions of a Dirac operator~\cite{ChamseddineConnes1997,CCvS2019}, and from determinantal or Eddington-inspired Born--Infeld gravity, which uses a square root of a determinant rather than its logarithm~\cite{DeserGibbons1998,BanadosFerreira2010,BeltranJimenez2018}. Although the spectrum of general $\mathcal L(g,R_{\mu\nu\rho\sigma})$ theories is known~\cite{Gullu2010,BuenoCano2016,Bueno2017}, we derive it directly for this family. The ghost-free $q=0$ branch also connects to nonminimally coupled $f(R,\mathcal L_m)$ gravity~\cite{Bertolami2007,HarkoLobo2010}.

Section~\ref{sec:gfe} reviews GfE; unless otherwise noted, those results are from Ref.~\cite{GfE}. Sections~\ref{sec:struct}--\ref{sec:Lambda} develop the action's structural properties, define the entropy families, derive their spectra, and analyze the G-field, ghost-free branches, and volume-constraint cosmological constant. Section~\ref{sec:disc} summarizes the results. The appendices provide derivations and report symbolic and numerical checks.

\paragraph*{Conventions.} The signature is $(-,+,\dots,+)$ and $\hbar=c=1$. The curvature conventions are
\begin{gather}
R^\rho{}_{\sigma\mu\nu}=\partial_\mu\Gamma^\rho_{\nu\sigma}-\partial_\nu\Gamma^\rho_{\mu\sigma}+\Gamma^\rho_{\mu\lambda}\Gamma^\lambda_{\nu\sigma}-\Gamma^\rho_{\nu\lambda}\Gamma^\lambda_{\mu\sigma},\\
R_{\mu\nu}=R^\rho{}_{\mu\rho\nu},\qquad R=g^{\mu\nu}R_{\mu\nu},
\end{gather}
so that de Sitter space has $R>0$, and $\Box:=g^{\mu\nu}\nabla_\mu\nabla_\nu$, so that a mode of mass $m_s$ obeys $(\Box-m_s^2)h=0$. The spacetime dimension is $d$, and numbers are given for $d=4$, with $G$ Newton's constant, $\lP:=\sqrt G$ the Planck length (in $d$ dimensions $\lP^{d-2}:=G$) and $M_P:=\lP^{-1}$ the Planck mass. Coordinates carry dimensions of length and the metric is dimensionless; the Lagrangian $\mathcal L$ is dimensionless, the couplings $\alpha$ and $\beta$ have dimensions of length$^d$ and length$^2$, and the Dirac--K\"ahler fields have the canonical mass dimension $(d-2)/2$. $G_{\mu\nu}$ is the Einstein tensor, $C_{\mu\nu\rho\sigma}$ the Weyl tensor, $C^2:=C_{\mu\nu\rho\sigma}C^{\mu\nu\rho\sigma}$, and $\mathscr K:=R_{\mu\nu\rho\sigma}R^{\mu\nu\rho\sigma}$ the Kretschmann scalar. The symbol $\mathcal R$ stands for a typical curvature component, so that $O(\mathcal R^n)$ denotes terms of $n$th order in the curvature. A few letters carry two standard meanings, distinguished by context: $F$ is the generating function of Sec.~\ref{sec:family} and $F_{\mu\nu}$ the field strength of the 1-form; $\rho$ is a density matrix in Secs.~\ref{sec:gfe}--\ref{sec:family} and an energy density elsewhere; $M$ is a mass, whereas $\bM$ and $\tM$ are matter operators.

\section{Gravity from entropy: a brief review}\label{sec:gfe}
This section summarizes Ref.~\cite{GfE}; unless stated otherwise, the definitions and assumptions follow that work. The hypotheses used below are labelled \rev{(A1)--(A6)}.

\subsection{Metrics as quantum operators}\label{sec:ops}
\begin{hypothesis}{Geometry}{as:geom}Spacetime is a $d$-dimensional manifold with Lorentzian metric $g_{\mu\nu}$ and the torsion-free, metric-compatible (Levi-Civita) connection $\nabla$.
\end{hypothesis}
The eigenvalues of a covariant rank-2 tensor $\hat G_{\mu\nu}$ are defined by $\hat G_{\mu\nu}e^\nu=\lambda e_\mu$ or, equivalently, as the eigenvalues of the \emph{mixed} tensor
\begin{equation}
\hat N_\mu{}^\nu:=\hat G_{\mu\sigma}g^{\sigma\nu}.
\label{eq:mixed}
\end{equation}
Under a coordinate transformation with Jacobian $J$, the mixed tensor transforms by similarity, $\hat N\to J^{T}\hat N(J^{T})^{-1}$. Its characteristic polynomial and spectrum are therefore invariant. By contrast, the covariant tensor transforms by congruence, $\hat G\to J^T\hat GJ$, which does not preserve eigenvalues. Thus the eigenvalues defined through $\hat N$ are diffeomorphism-invariant scalars. For $\hat G=g$, $\hat N=\delta_\mu^\nu$, and every eigenvalue is unity.

Functions of a metric are defined through $\hat N$:
\begin{gather}
[f(\hat G)]_{\mu\nu}:=[f(\hat N)]_\mu{}^\rho g_{\rho\nu},\qquad [\hat G^{-1}]^{\mu\nu}:=g^{\mu\rho}[\hat N^{-1}]_\rho{}^\nu,\\
\Tr\hat G:=\TrM\hat N=\hat G_{\mu\nu}g^{\mu\nu},
\label{eq:funcs}
\end{gather}
where $\TrM$ is the ordinary matrix trace. These reproduce the spectral definitions of Ref.~\cite{GfE} whenever $\hat N$ is diagonalizable. The corresponding entropy and cross-entropy are
\begin{align}
H_{\rm vN}(\hat G)&=\Tr\hat G\ln\hat G^{-1}=-\TrM\hat N\ln\hat N=-\sum_\lambda\lambda\ln\lambda,\label{eq:H}\\
\Tr g\ln\hat G^{-1}&=-\TrM\ln\hat N=-\sum_\lambda\ln\lambda,\label{eq:cross}
\end{align}
so that the von Neumann entropy $H_{\rm vN}(g)$ vanishes identically.

\begin{hypothesis}{Metrics as quantum operators}{as:ops}Metrics are interpreted as quantum operators, or the \emph{renormalizable effective density matrices} of Ref.~\cite{GfE}. \rev{Reference~\cite{GfE} requires the metrics to be invertible but not to have unit trace; we read ``renormalizable'' in this sense,} not \rev{as} renormalizability in quantum field theory. The metrics are invertible, so $\hat G$ and $\hat G^{-1}$ enter on the same footing, as required by Lorentz invariance; their traces need not be unity. Positivity is imposed by requiring the spectrum of $\hat N=\hat Gg^{-1}$ [Eq.~\eqref{eq:mixed}] to be real and positive. This condition is needed to define the logarithm and noninteger powers, and we impose it for every member of the family in Sec.~\ref{sec:family}, including those with polynomial generating functions.
\end{hypothesis}

In operator-algebraic quantum field theory~\cite{Araki1999,Araki1976,OhyaPetz,Witten2018}, relative entropy is defined for positive operators that need not be normalized density matrices. GfE applies this broader notion to geometry; in particular, normalizing a metric would not generally normalize its inverse or dual.

\subsection{Warm-up: a scalar field}\label{sec:warm}
For a complex scalar $\phi$, Ref.~\cite{GfE} defines the induced metric
\begin{equation}
\bG=g+\alpha\bM,\qquad M_{\mu\nu}=\nabla_\mu\bar\phi\,\nabla_\nu\phi,
\label{eq:warmG}
\end{equation}
with $\alpha=\alpha'\lP^d$ ($\alpha'$ dimensionless). For real $\phi$ on a Euclidean flat background, $\bG$ is the first fundamental form of the graph $\{(x,\phi(x))\}\subset\mathbb R^{d+1}$~\cite{KeskiVakkuri2022}; it encodes how the field bends the geometry. Define $u_\mu=\nabla_\mu\bar\phi$, $v^\nu=\nabla^\nu\phi$, and $w=|\nabla\phi|^2=v^\mu u_\mu$. The mixed operator $A=\bM g^{-1}=uv^T$ has rank one and satisfies
\begin{gather}
A^n=w^{n-1}A,\nonumber\\ f(I+\alpha A)=f(1)I+\frac{f(1+\alpha w)-f(1)}{w}A
\label{eq:rank1}
\end{gather}
for every analytic $f$; the second identity follows by expanding $f$ in a power series and using $A^n=w^{n-1}A$. The spectrum of $\mathbf N=\bG g^{-1}$ is $\{1^{(d-1)},1+\alpha w\}$. The relative entropy of $g$ with respect to $\bG$ therefore gives the Lagrangian
\begin{equation}
\mathcal L=-\Tr g\ln g^{-1}+\Tr g\ln\bG^{-1}=-\ln(1+\alpha|\nabla\phi|^2),
\label{eq:Lwarm}
\end{equation}
which reduces to the massless Klein--Gordon Lagrangian $-\alpha|\nabla\phi|^2$ for $\alpha w\ll1$. Varying $\phi$ gives $\nabla_\mu[j(w)\nabla^\mu\phi]=0$, where $j=\alpha/(1+\alpha w)$; varying the metric gives
\begin{equation}
\frac{1}{\sg}\frac{\delta(\sg\mathcal L)}{\delta g^{\mu\nu}}=-j(w)M_{(\mu\nu)}-\tfrac12\mathcal L\,g_{\mu\nu}=0,
\label{eq:warmmetric}
\end{equation}
where we used $\delta\sg=-\frac12\sg\,g_{\mu\nu}\delta g^{\mu\nu}$. This scalar model has no mass term and leaves $g$ undetermined in vacuum. Moreover, for $d\ge3$, writing $\nabla\phi=a+ib$ with real $a,b$ shows that $M_{(\mu\nu)}=a_\mu a_\nu+b_\mu b_\nu$ has rank at most two. Equation~\eqref{eq:warmmetric} then forces $\mathcal L=0$ and $\nabla\phi=0$. The construction below addresses both limitations.

\subsection{Topological fields and topological metrics}\label{sec:topo}
\begin{hypothesis}{Topological matter}{as:matter}Matter is a complex Dirac--K\"ahler boson~\cite{Kahler1962,BecherJoos1982,Kruglov2002,Berenstein2023},
\begin{equation}
\ket\Phi=\phi\oplus\omega_\mu dx^\mu\oplus\zeta_{\mu\nu}dx^\mu\wedge dx^\nu,\qquad \zeta_{\mu\nu}=-\zeta_{\nu\mu},
\end{equation}
truncated at 2-forms, as in Ref.~\cite{GfE}.
\end{hypothesis}
The metric induces the inner products $\langle dx^\mu,dx^\nu\rangle=g^{\mu\nu}$ and $\langle dx^\rho\wedge dx^\sigma,dx^\eta\wedge dx^\theta\rangle=g^{\rho\eta}g^{\sigma\theta}-g^{\rho\theta}g^{\sigma\eta}$. In the sum $\zeta_{\mu\nu}dx^\mu\wedge dx^\nu$, each independent component appears twice, as in Ref.~\cite{GfE}. The factor $\tfrac12$ in the metric on 2-forms compensates for this:
\begin{equation}
[g_{(2)}]_{\mu\nu\rho\sigma}=\tfrac12(g_{\mu\rho}g_{\nu\sigma}-g_{\mu\sigma}g_{\nu\rho}).
\label{eq:g2}
\end{equation}
The topological metric is $\tg=1\oplus g\oplus g_{(2)}$. A metric $\bG_{(2)}$ on 2-forms acts as $\zeta_{\mu\nu}\mapsto[G_{(2)}]_{\mu\nu\rho\sigma}[g_{(2)}]^{\rho\sigma\eta\theta}\zeta_{\eta\theta}$. Composition uses a full index sum, and the identity is
\begin{equation}
[g_{(2)}]_{\mu\nu\rho\sigma}[g_{(2)}]^{\rho\sigma\eta\theta}=\tfrac12(\delta_\mu^\eta\delta_\nu^\theta-\delta_\mu^\theta\delta_\nu^\eta)=:\mathbf 1_{\mu\nu}{}^{\eta\theta}.
\end{equation}
The identity on the fibre is $\tI:=\tg\tg^{-1}=1\oplus\delta_\mu{}^\nu\oplus\mathbf 1_{\mu\nu}{}^{\rho\sigma}$. Thus every eigenvalue of $g_{(2)}$ is unity, and the unrestricted trace is $\TrM g_{(2)}g^{-1}_{(2)}=\mathbf 1_{\mu\nu}{}^{\mu\nu}=\binom d2$. We denote traces within a block by $\TrM$ and the trace over all form sectors by $\TrF$. The local fibre of topological fields consequently has dimension
\begin{equation}
\cN:=\TrF\tI=1+d+\binom d2\overset{d=4}{=}11 .
\label{eq:calN}
\end{equation}

The Dirac--K\"ahler operator $\mathbb D=d+\delta$, restricted to degrees $\le2$ (the 3-form $d\zeta$ \rev{does not appear in Ref.~\cite{GfE}}), acts as
\begin{equation}
\mathbb D\ket\Phi=-\nabla^\mu\omega_\mu\oplus\Pi_\mu dx^\mu\oplus\tfrac12F_{\mu\nu}dx^\mu\wedge dx^\nu,
\label{eq:DK}
\end{equation}
where $\Pi_\mu=\nabla_\mu\phi-\nabla^\rho\zeta_{\rho\mu}$ and $F_{\mu\nu}=\nabla_\mu\omega_\nu-\nabla_\nu\omega_\mu$, so that $\tfrac12F_{\mu\nu}=\nabla_{[\mu}\omega_{\nu]}$. On scalars $\mathbb D^2\phi=\delta d\phi=-\Box\phi$. We keep the normalization of Ref.~\cite{GfE}. Because the sum over $\mu,\nu$ in the 2-form includes both index orderings, the 2-form contribution to $\Pi_\mu$ may differ by a factor of 2 from other codifferential conventions. This affects only the relative normalization of the 2-form kinetic terms and none of our results. The bra in $\ket\Phi\bra\Phi$ denotes complex conjugation; for example, $[\ket\Phi\bra\Phi]_{(1)\mu\nu}=\bar\omega_\mu\omega_\nu$.

\begin{hypothesis}{Induced metric}{as:induced}The metric induced by matter and curvature is~\cite{GfE}
\begin{gather}
\tG=\tg+\alpha\tM-\beta\tR,\label{eq:tG}\\
\tM=\mathbb D\ket\Phi\bra\Phi\mathbb D+(m^2+\xi R)\ket\Phi\bra\Phi,\\
\tR=R\oplus R_{\mu\nu}\,dx^\mu\otimes dx^\nu\nonumber\\ \oplus R_{\mu\nu\rho\sigma}(dx^\mu\wedge dx^\nu)\otimes(dx^\rho\wedge dx^\sigma),
\end{gather}
where $m$ is the bare mass of $\Phi$ and $\xi$ a dimensionless non-minimal coupling, with $\alpha=\alpha'\lP^d>0$ and $\beta=\beta'\lP^2>0$, so that $\tG$ is dimensionless.
\end{hypothesis}
In the density-matrix interpretation, $\mathbb D\ket\Phi\bra\Phi\mathbb D$ is a gradient-built pure state that generates the kinetic terms. The projector-like term $(m^2+\xi R)\ket\Phi\bra\Phi$ generates the mass and nonminimal coupling, with conformal coupling $\xi=(d-2)/[4(d-1)]$. The curvature contribution $-\beta\tR$ is not a sum of projectors and makes the metric dynamical even in vacuum. The Riemann tensor belongs in the 2-form block because its symmetries make it a Hermitian metric on bivectors: $R_{\mu\nu\rho\sigma}=-R_{\nu\mu\rho\sigma}=-R_{\mu\nu\sigma\rho}=R_{\rho\sigma\mu\nu}$.

Sector by sector,
\begin{align}
M_{(0)}&=|\nabla^\mu\omega_\mu|^2+(m^2+\xi R)|\phi|^2,\nonumber\\
[M_{(1)}]_{\mu\nu}&=\bar\Pi_\mu\Pi_\nu+(m^2+\xi R)\bar\omega_\mu\omega_\nu,\label{eq:Mblocks}\\
[M_{(2)}]_{\mu\nu\rho\sigma}&=\tfrac14\bar F_{\mu\nu}F_{\rho\sigma}+(m^2+\xi R)\bar\zeta_{\mu\nu}\zeta_{\rho\sigma}.\nonumber
\end{align}
The central object of this paper is the mixed operator
\begin{equation}
\tN:=\tG\tg^{-1}=\tI+\alpha\tM\tg^{-1}-\beta\tR\tg^{-1},
\label{eq:tN}
\end{equation}
whose blocks are the matrices $\mathbf N_{(n)}$ of Ref.~\cite{GfE}; the curvature operator $\tX:=\tR\tg^{-1}$ has blocks $\mathbf X_{(n)}$. The fibre trace $\TrF$ sums the ordinary traces of the three blocks, using unrestricted index sums in the 2-form block. Operators in that block act on antisymmetric index pairs, and their functions are defined on this subspace, where $\mathbf 1_{\mu\nu}{}^{\rho\sigma}$ is the identity.

\paragraph*{Curvature traces.} Each block of $\tX$ has trace $R$, giving $\TrF\tX=3R$. The traces of the squared blocks give the three quadratic curvature invariants:
\begin{equation}
\TrF\tX^2=R^2+R_{\mu\nu}R^{\mu\nu}+R_{\mu\nu\rho\sigma}R^{\mu\nu\rho\sigma}.
\label{eq:TrX2}
\end{equation}
This is immediate for the 0- and 1-form blocks; the 2-form block is treated in Appendix~\ref{app:weyl}.

\subsection{\texorpdfstring{\rev{The GfE action}}{The GfE action}}\label{sec:GfEaction}
\rev{The Gravity from Entropy action, or GfE action for short, is}~\cite{GfE}
\begin{gather}
\mathcal L=-\TrF\tg\ln\tg^{-1}+\TrF\tg\ln\tG^{-1}=-\TrF\ln\tN,\label{eq:L}\\
\mathcal S=\frac1{\lP^d}\int\sg\,\mathcal L\,\dr .
\end{gather}
\rev{Reference~\cite{GfE} introduced $\mathcal L$ as the quantum relative entropy of $\tg$ with respect to $\tG$; Refs.~\cite{GfEthermo,BianconiLambda2026} call it the geometric quantum relative entropy (GQRE).} The first term vanishes because all eigenvalues of $\tg$ equal one. \rev{The GQRE goes beyond the standard quantum relative entropy of states: by assumption~(\ref{as:ops}), the metric itself is treated as a quantum operator. This is the central physical insight of GfE, and the constructions of this paper build on it.}
\paragraph*{Relation to Araki's relative entropy.} \rev{Reference~\cite{GfE} (Appendix~C there) relates the GfE Lagrangian to Araki's relative entropy; we restate the relation in the present notation.} In the Hilbert--Schmidt representation, the relative entropy of positive $\rho,\sigma$ is~\cite{Araki1976}
\begin{equation*}
S(\rho\Vert\sigma)=-\langle\rho^{1/2},\ln\Delta_{\sigma|\rho}\,\rho^{1/2}\rangle=\Tr\rho\ln\rho-\Tr\rho\ln\sigma,
\end{equation*}
where $\rho^{1/2}$ is the vector representative of $\rho$ and $\Delta_{\sigma|\rho}X=\sigma X\rho^{-1}$ is the relative modular operator. In the GfE fibre algebra, products are defined using $\tg^{-1}$; thus $\tg$ represents the identity, $\rho^{1/2}=\tI$, and $\Delta$ acts by left multiplication by $\tN$. This gives $S(\tg\Vert\tG)=-\TrF\ln\tN$. Reference~\cite{GfE} denotes $\tG\tg^{-1}$ by $\bm\Delta^{1/2}_{\tG,g}$; this difference in modular-operator convention does not change the resulting Lagrangian. \rev{The GQRE thus has the algebraic form of Araki's relative entropy, but its arguments are metrics rather than states.}

\subsection{Weak-coupling limit and Newton's constant}\label{sec:weak}
\begin{hypothesis}{Weak coupling}{as:weak}Wherever series are used, $\|\tN-\tI\|\ll1$. With $\beta=\beta'\lP^2$, this means curvature radii $\gg\lP$ and energy densities well below the Planck density.
\end{hypothesis}
Expanding $-\ln(\tI+Y)=-Y+\frac12Y^2-\dots$ with $Y=\alpha\tM\tg^{-1}-\beta\tX$ and using $\TrF\tX=3R$, we find
\begin{equation}
\mathcal L=3\beta R-\alpha\big[\bra\Phi\mathbb D\tg^{-1}\mathbb D\ket\Phi+(m^2+\xi R)\langle\Phi|\Phi\rangle\big]+O(Y^2),
\label{eq:EH}
\end{equation}
with $\langle\Phi|\Phi\rangle=|\phi|^2+\bar\omega^\mu\omega_\mu+\bar\zeta^{\mu\nu}\zeta_{\mu\nu}$ and
\begin{align}
\bra\Phi\mathbb D\tg^{-1}\mathbb D\ket\Phi&=|\nabla\phi|^2+|\nabla^\mu\omega_\mu|^2+|\nabla^\rho\zeta_{\rho\mu}|^2\nonumber\\
&\quad+\tfrac14\bar F^{\mu\nu}F_{\mu\nu}-2\,\mathrm{Re}(\nabla^\mu\bar\phi\,\nabla^\rho\zeta_{\rho\mu}).
\end{align}
The cross term is a total derivative: after integration by parts, the symmetry of $\nabla^\rho\nabla^\mu\bar\phi$ and the antisymmetry of $\zeta_{\rho\mu}$ make it vanish. The 2-form block supplies the Maxwell-like kinetic term $\tfrac14\bar F^{\mu\nu}F_{\mu\nu}$ for the 1-form $\omega$. Define $\mathcal L_m:=-\TrF\tM\tg^{-1}$, the negative of the bracket in Eq.~\eqref{eq:EH}. The linear term is then $3\beta R+\alpha\mathcal L_m$, with $\lP^{-d}\alpha\mathcal L_m=\alpha'\mathcal L_m$ in the action.
For $\omega=\zeta=0$, the leading action is Einstein gravity coupled to a Klein--Gordon field with nonminimal coupling $\xi R$~\cite{GfE,MukhanovWinitzki}. Comparing with $\int\sg[R/(16\pi G)+\mathcal L_m]$ in $d=4$, using the same $\mathcal L_m$, fixes
\begin{equation}
\alpha'=1,\qquad \frac{3\beta'}{\lP^2}=\frac1{16\pi G}\ \Rightarrow\ \beta'=\frac{1}{48\pi}\ \ (\lP^2=G).
\label{eq:norm}
\end{equation}
At this order the metric equation is $G_{\mu\nu}=8\pi GT_{\mu\nu}$, where $T_{\mu\nu}$ is the stress tensor of $\mathcal L_m$ and $8\pi G=\alpha/(6\beta)$.

\subsection{G-field formulation}\label{sec:GfE-Gfield}
Since the Lagrangian~\eqref{eq:L} depends on $\tG$ only through $\tN$, Ref.~\cite{GfE} introduces an auxiliary operator field $\tTh$, with the same block and index structure as $\tN$, and imposes $\tN=\tTh$ with Lagrange multipliers $\tGc$, the G-field:
\begin{equation}
\tilde{\mathcal L}=-\TrF\ln\tTh-\TrF\tGc(\tN-\tTh).
\label{eq:Ltilde}
\end{equation}
All components are mixed tensors, e.g.\ $[\bm{\mathcal G}_{(1)}]_\mu{}^\nu$ and $[\bm{\mathcal G}_{(2)}]_{\mu\nu}{}^{\rho\sigma}$. Using $\delta\TrF\ln\tTh=\TrF\tTh^{-1}\delta\tTh$, the auxiliary equations are
\begin{equation}
\delta_{\tGc}:\ \tTh=\tN,\qquad \delta_{\tTh}:\ \tGc=\tTh^{-1}.
\label{eq:aux}
\end{equation}
Substituting $\tTh=\tGc^{-1}$ gives
\begin{align}
\tilde{\mathcal L}&=\TrF\ln\tGc-\TrF\tGc(\tI+\alpha\tM\tg^{-1}-\beta\tX)+\cN\nonumber\\
&=\beta\big(\RG-2\LG\big)-\alpha\MG,\label{eq:dressed}
\end{align}
where
\begin{gather}
\RG=\TrF\tg_{\mathcal G}^{-1}\tR,\qquad \MG=\TrF\tg_{\mathcal G}^{-1}\tM,\\
\LG=\frac1{2\beta}\TrF(\tGc-\tI-\ln\tGc),\label{eq:LG}
\end{gather}
and $\tg_{\mathcal G}:=\tGc^{-1}\tg$ is the \emph{dressed metric}. We keep the normalization of $\LG$ of Refs.~\cite{GfE,GfEthermo}. \rev{Near $\tGc=\tI$, $\beta(\RG-2\LG)\simeq3\beta(R-2\LG/3)$, so $\LG/3$ enters like a standard cosmological constant.}
\begin{hypothesis}{Physical G-field}{as:Gfield}\rev{Following Ref.~\cite{GfE},} the multipliers $\tGc$ are promoted to physical fields, in the same way that the Lagrange multipliers of statistical mechanics (temperature, chemical potential) are measurable. In metric variations the mixed components of $\tGc$ are held fixed\rev{; Ref.~\cite{GfE} does not state this choice explicitly, but it is consistent with the mixed-index form of its field equations}. This hypothesis is interpretive: classically it adds no dynamics, since the G-field equations are algebraic (Sec.~\ref{sec:equiv}).
\end{hypothesis}
Equation~\eqref{eq:dressed} is a dressed Einstein--Hilbert action: curvature and matter couple through the dressed metric, while $\LG$ appears as a cosmological term. The potential $\LG$ is non-negative, vanishes at $\tGc=\tI$, and is quadratic near that point. We identify it with the Lindblad divergence $D(\tI\Vert\tGc)$ in Sec.~\ref{sec:equiv}. Setting $\tGc=\tI$ gives $\RG=3R$ and recovers the Einstein term.

\section{Structural properties of gravity from entropy}\label{sec:struct}
\subsection{\texorpdfstring{\rev{The GfE Lagrangian as a logarithmic determinant}}{The GfE Lagrangian as a logarithmic determinant}}\label{sec:logdet}
On the positive-spectrum domain assumed here, $\TrM\ln A=\ln\det A$: both sides equal the sum of logarithms of the eigenvalues, counted with algebraic multiplicity. This remains true when $A$ is not diagonalizable. Applied block by block to Eq.~\eqref{eq:L}, it gives
\begin{equation}
\mathcal L=-\ln\det\tN=\revm{-\sum_{n=0}^2\ln\det\mathbf N_{(n)}} .
\label{eq:logdet}
\end{equation}
\rev{Here $\det\mathbf N_{(n)}$ is the determinant of the mixed operator $\mathbf N_{(n)}=\bG_{(n)}g_{(n)}^{-1}$, i.e.\ the product of its eigenvalues, and is a scalar. It should not be decomposed as $\det\bG_{(n)}/\det g_{(n)}$: the determinant of the metric is defined through the tetrads, $|\det g|=(\det e)^2$, whereas there is no corresponding notion of determinant for the induced operator $\tG$. With $W:=\prod_n\det(g_{(n)}\bG_{(n)}^{-1})=(\det\tN)^{-1}$, as in Ref.~\cite{GfEthermo}, Eq.~\eqref{eq:logdet} takes the form $\mathcal L=\ln W$. Reference~\cite{GfEthermo} interprets it as a Gibbs--Maxwell entropy, with $W$ quantifying the degrees of freedom of $\tG$ that can be encoded by $\tg$. In GfE the Lagrangian is therefore at once a quantum relative entropy, the GQRE, and an entropy of Boltzmann--Gibbs form, and the action can be interpreted as the entropy of the geometric degrees of freedom of spacetime~\cite{GfEthermo}. This statistical-mechanical interpretation rests on the logarithm and is specific to GfE. For the generalized entropies of Sec.~\ref{sec:family} the Lagrangian is no longer the logarithm of a determinant: $-\TrF\ln_q\tN$ differs from $-\ln_q\det\tN$ for $q\ne1$, since $\ln_q(xy)=\ln_qx+\ln_qy+(1-q)\ln_qx\ln_qy$, and the R\'enyi and Sharma--Mittal Lagrangians depend on the power sum $\TrF\tN^{1-q}$ rather than on $\det\tN$. The generalizations retain the relative-entropy structure but lose this Boltzmann--Gibbs interpretation.} The Dirac--Born--Infeld (DBI) Lagrangian~\cite{BornInfeld1934,Dirac1962}, by contrast, uses \rev{the square root of a determinant rather than its logarithm}. In the warm-up model, the matrix determinant lemma gives $\revm{\det\mathbf N=1+\alpha w}$, recovering Eq.~\eqref{eq:Lwarm}.

\subsection{Origin of the Einstein--Hilbert term}
For positive, not necessarily normalized operators, the Lindblad divergence is the non-negative functional~\cite{Lindblad1973,OhyaPetz}
\begin{equation}
D(\rho\Vert\sigma)=\Tr\big[\rho\ln\rho-\rho\ln\sigma-\rho+\sigma\big]\ \ge0 .
\label{eq:lindblad}
\end{equation}
For $\rho=\tI$, it is $\sum_\lambda(\lambda-1-\ln\lambda)$, whose terms are non-negative by $\ln x\le x-1$. The GfE Lagrangian can therefore be written
\begin{equation}
\mathcal L=D(\tI\Vert\tN)-\TrF(\tN-\tI),
\label{eq:split}
\end{equation}
where $D(\tI\Vert\tN)=O((\tN-\tI)^2)$, whereas $-\TrF(\tN-\tI)=3\beta R-\alpha\TrF\tM\tg^{-1}$ exactly. The Einstein--Hilbert term is therefore the linear, sign-indefinite contribution, while the positive divergence begins at second order and generates the higher-curvature terms. Because the linear term depends only on $\TrF\tX=3R$, it is shared by every family member $-\TrF F(\tN)$ with $F(1)=0$ and $F'(1)=1$. The information-theoretic part first affects the action through $F''(1)$ and higher derivatives. Thus $\mathcal L$ itself is not bounded below. In what follows, ``relative entropy'' denotes the unnormalized Umegaki functional $\mathcal L=-\TrF\ln\tN$, while $D$ denotes the non-negative Lindblad divergence.

\subsection{Lorentzian spectral domain}\label{sec:lorentz}
The operator $\tN$ is self-adjoint with respect to the \emph{indefinite} form $\tg$, so the field space is a Krein space~\cite{Bognar1974}. Unlike a self-adjoint operator on a positive-definite space, it can have complex eigenvalues and Jordan blocks. For example, in $d=2$, the tensor $\hat G=g+b(dx^0\otimes dx^1+dx^1\otimes dx^0)$ gives
\begin{equation*}
\hat N=\begin{pmatrix}1&b\\-b&1\end{pmatrix},
\end{equation*}
with eigenvalues $1\pm ib$, analogous to the Segre type $[Z\bar Z11]$ of the Ricci operator~\cite{Stephani2003}. For a null gradient in Eq.~\eqref{eq:warmG}, for instance, $A^2=0\ne A$ and $\ln\mathbf N=\alpha A$. A positive-definite (Euclidean) $\hat G$ also fails the positivity condition in assumption~(\ref{as:ops}): $\hat Gg^{-1}$ is similar to $\hat G^{1/2}g^{-1}\hat G^{1/2}$, which has one negative eigenvalue by Sylvester's law of inertia.

The complex-spectrum and negative-eigenvalue examples violate assumption~(\ref{as:ops}); the null-gradient example does not, since its eigenvalues are all unity. For the complex-conjugate pair, the principal-branch trace logarithm remains real. In the null-gradient case, the logarithm is instead defined by the nilpotent expansion $\ln(\mathbf I+\alpha A)=\alpha A$. Negative eigenvalues make $\mathcal L$ complex. The backgrounds analyzed here have diagonalizable, real spectra, for which the condition reduces to positivity of the eigenvalues.

\subsection{G-field: equivalence, emergent cosmological term and Wald tensor}\label{sec:equiv}
The original and auxiliary Lagrangians, Eqs.~\eqref{eq:L} and~\eqref{eq:Ltilde}, are classically equivalent because the auxiliary equations~\eqref{eq:aux} are algebraic and eliminate $\tTh$ and $\tGc$. On shell, the dressed metric equals the induced metric, $\tg_{\mathcal G}=\tGc^{-1}\tg=\tN\tg=\tG$. Equations~\eqref{eq:LG} and~\eqref{eq:lindblad} also identify the emergent term as the Lindblad divergence $\LG=\frac1{2\beta}D(\tI\Vert\tGc)\ge0$. In vacuum, $\tGc=(\tI-\beta\tX)^{-1}=\tI+\beta\tX+\beta^2\tX^2+\cdots$. Expanding $x-1-\ln x=\frac12(x-1)^2+O((x-1)^3)$ gives $\LG=\frac{\beta}{4}\TrF\tX^2+\cdots$, and hence, by Eq.~\eqref{eq:TrX2},
\begin{equation}
\LG=\frac\beta4\big(R^2+R_{\mu\nu}R^{\mu\nu}+R_{\mu\nu\rho\sigma}R^{\mu\nu\rho\sigma}\big)+O(\beta^2\mathcal R^3).
\label{eq:LGvac}
\end{equation}
Matter propagates in the metric it induces, while $\tg$ supplies the volume measure $\sg$. \rev{In Ref.~\cite{GfE}, $\LG$ is the emergent cosmological constant of the theory, small and positive near $\tGc=\tI$, and Refs.~\cite{GfEthermo,BianconiLambda2026} identify the associated energy density with a dynamical dark-energy contribution. On shell, however,} it does not act as a constant cosmological term. In vacuum, $\tGc=\tI+\beta\tX+O(\beta^2\mathcal R^2)$, which gives
\begin{gather*}
\beta\RG=3\beta R+4\beta\LG+O(\beta^3\mathcal R^3),\\
\beta(\RG-2\LG)=3\beta\big(R+\tfrac23\LG\big)+O(\beta^3\mathcal R^3),
\end{gather*}
These relations show that the $\LG$-dependent part of the dressed Lagrangian contributes at curvature-squared order, as in Eq.~\eqref{eq:quadGfE}. In $d=4$, this term does not affect the vacuum equations on maximally symmetric spaces~\cite{DeserTekin2003,PaperII}.

\paragraph*{G-field as Wald tensor.} The G-field also has a geometric interpretation. For a Lagrangian depending on the metric and Riemann tensor, $P^{\mu\nu\rho\sigma}:=\partial\mathcal L/\partial R_{\mu\nu\rho\sigma}$ enters both the field equations~\cite{Padmanabhan2013} and the Wald entropy of stationary horizons~\cite{Wald1993,IyerWald1994}. In GfE, curvature enters through $\tN$ and, in vacuum (or more generally for $\xi=0$), only through $-\beta\tX$. The chain rule and the G-field equation imply $\partial\mathcal L/\partial\tX=\beta\tGc$ block by block. On shell [Eq.~\eqref{eq:PG} of Appendix~\ref{app:var}], this gives
\begin{equation*}
P^{\mu\nu\rho\sigma}=\beta\,\mathbb P\Big[\bm{\mathcal G}_{(0)}\,g^{\mu\rho}g^{\nu\sigma}+g^{\mu\rho}\bm{\mathcal G}_{(1)}{}^{\nu\sigma}+\bm{\mathcal G}_{(2)}{}^{\mu\nu\rho\sigma}\Big],
\end{equation*}
where $\mathbb P$ projects onto tensors with the algebraic symmetries of the Riemann tensor. For the action $\lP^{-d}\int\sg\,\mathcal L\,\dr$, the Wald entropy~\cite{Wald1993,IyerWald1994} of a bifurcate Killing horizon with bifurcation surface $\mathcal H$ and binormal $n_{\mu\nu}$, normalized by $n_{\mu\nu}n^{\mu\nu}=-2$, is $S_W=-2\pi\lP^{-d}\oint_{\mathcal H}P^{\mu\nu\rho\sigma}n_{\mu\nu}n_{\rho\sigma}$. It is therefore determined by the horizon integral of the binormal contraction of the G-field, with normalization fixed by $\beta$ and $\lP$.

\section{\texorpdfstring{\rev{Gravity from generalized entropy (GfGE)}}{Gravity from generalized entropy (GfGE)}}\label{sec:family}
\subsection{Definition}
\rev{To distinguish them from the original GfE theory, we refer to actions and Lagrangians built from generalized entropies as gravity from generalized entropy (GfGE).} We consider the family
\begin{equation}
\mathcal L_{F,\Psi}=\Psi\big(\mathcal L_F\big),\qquad \mathcal L_F:=-\TrF F(\tN),
\label{eq:family}
\end{equation}
where $F$ is analytic near 1, with $F(1)=0$ and $F'(1)=1$, and $\Psi$ is analytic near 0, with $\Psi(0)=0$ and $\Psi'(0)=1$. These normalizations preserve the Einstein limit (Sec.~\ref{sec:spectrum}). Define $\kappa_2:=-F''(1)/2$ and $\psi_2:=\Psi''(0)$. We also allow sector-dependent couplings and general curvature blocks:
\begin{equation}
\tN=\tI+\bigoplus_{n=0}^{2}\big(\alpha_n\bM_{(n)}g_{(n)}^{-1}-\beta_n\mathbf X_{(n)}\big).
\label{eq:Ngen}
\end{equation}
The blocks $\mathbf X_{(n)}$ are linear in the curvature, and to characterize them we use the $d=4$ decomposition of the curvature operator on bivectors, $R_{\mu\nu}{}^{\rho\sigma}=\mathsf C+\mathsf S+\mathsf R_0$, into its Weyl, traceless-Ricci and scalar parts (Appendix~\ref{app:weyl}). With $S_{\mu\nu}=R_{\mu\nu}-\frac14Rg_{\mu\nu}$, the most general blocks are
\begin{gather}
\mathbf X_{(0)}=\gamma_{R0}R,\qquad \mathbf X_{(1)}=\gamma_{S1}S_\mu{}^\nu+\gamma_{R1}\tfrac R4\delta_\mu^\nu,\label{eq:blocks}\\
\mathbf X_{(2)}=\gamma_W\mathsf C+\gamma_{S2}\mathsf S+\gamma_{R2}\mathsf R_0 .\nonumber
\end{gather}
GfE corresponds to $F=\ln$, $\Psi=\mathrm{id}$, all $\gamma=1$, $\alpha_n=\alpha$ and $\beta_n=\beta$\rev{; all other members are GfGE theories}.

\subsection{Non-extensive divergences with an identity reference}\label{sec:div}
For positive $\rho,\sigma$ we use the unnormalized Tsallis~\cite{Tsallis1988,Abe2003,Furuichi2004}, R\'enyi (Petz)~\cite{Renyi1961,Petz1986} and Sharma--Mittal~\cite{SharmaMittal1975} divergences:
\begin{align}
D^T_q(\rho\Vert\sigma)&=\frac{\Tr\rho-\Tr\rho^q\sigma^{1-q}}{1-q},\\
D^R_q(\rho\Vert\sigma)&=\frac1{q-1}\ln\frac{\Tr\rho^q\sigma^{1-q}}{\Tr\rho},\\
D^{SM}_{q,r}(\rho\Vert\sigma)&=\frac{1}{r-1}\Big[\Big(\frac{\Tr\rho^q\sigma^{1-q}}{\Tr\rho}\Big)^{\frac{r-1}{q-1}}-1\Big].
\end{align}
The Sharma--Mittal divergence tends to $D^R_q$ as $r\to1$ and to $D^T_q/\Tr\rho$ at $r=q$. As $q,r\to1$, Tsallis tends to the Umegaki relative entropy~\cite{Umegaki1962}, while R\'enyi and Sharma--Mittal tend to it divided by $\Tr\rho$. For noncommuting operators, other orderings are possible; one example is the sandwiched R\'enyi divergence, defined using $\Tr(\sigma^{(1-q)/2q}\rho\,\sigma^{(1-q)/2q})^q$~\cite{MullerLennert2013,WildeWinterYang2014}.
\paragraph*{Identity reference.} In GfE, the reference metric $\tg$ is represented by $\tI$ (Sec.~\ref{sec:gfe}), so $D(\tg\Vert\tG)=D(\tI\Vert\tN)$. With this identity reference, the Petz and sandwiched orderings coincide and depend on $\sigma$ only through $\Tr\sigma^{1-q}$. Define $\ln_qx=(x^{1-q}-1)/(1-q)$ and $\bar Z_q:=\av{\tN^{1-q}}$, where $\av{A}=\cN^{-1}\TrF A$. The resulting Lagrangians are
\begin{align}
\mathcal L^T_q&:=D^T_q(\tg\Vert\tG)=-\TrF\ln_q\tN=\frac{\cN(1-\bar Z_q)}{1-q},\label{eq:LT}\\
\mathcal L^R_q&:=\cN D^R_q(\tg\Vert\tG)=\frac{\cN}{q-1}\ln\bar Z_q,\label{eq:LR}\\
\mathcal L^{SM}_{q,r}&:=\cN D^{SM}_{q,r}(\tg\Vert\tG)=\frac{\cN}{r-1}\big[\bar Z_q^{\,s}-1\big],\quad s:=\frac{1-r}{1-q}.\label{eq:LSM}
\end{align}
The factor $\cN$ makes all three actions reduce to GfE at $q=r=1$. \rev{For $q,r\ne1$ these GfGE Lagrangians are not logarithms of $\det\tN$, so the Boltzmann--Gibbs form of the GfE action (Sec.~\ref{sec:logdet}) does not carry over.} We retain the GfE ordering, with $\tg$ as the reference. Reversing the order changes the spectrum: $-S(\tG\Vert\tg)=-\TrF\tN\ln\tN$ corresponds to $F(x)=x\ln x$ and $\kappa_2=-\frac12$, giving a non-tachyonic spin-2 ghost and a tachyonic scalar by Eq.~\eqref{eq:masses}. The ordering used here instead gives a healthy scalar.

\paragraph*{Range of the indices.} The indices $q$ and $r$ are real, with $q=1$ and $r=1$ reached as limits. The identity $\Tr\rho^q\sigma^{1-q}/\Tr\rho=e^{(q-1)D^R_q}$ implies
\begin{gather}
\frac{D^T_q}{\Tr\rho}=\mathfrak f_q\big(D^R_q\big),\qquad D^{SM}_{q,r}=\mathfrak f_r\big(D^R_q\big),\label{eq:frenyi}\\
\mathfrak f_k(u):=\frac{e^{(k-1)u}-1}{k-1},\nonumber
\end{gather}
and likewise $\mathcal L^T_q=\cN\mathfrak f_q(\mathcal L^R_q/\cN)$ and $\mathcal L^{SM}_{q,r}=\cN\mathfrak f_r(\mathcal L^R_q/\cN)$. Since each $\mathfrak f_k$ is strictly increasing and vanishes at zero, Tsallis and Sharma--Mittal inherit non-negativity, faithfulness, and data processing from the corresponding R\'enyi divergence. These properties depend on $q$, not $r$, which is unrestricted in the original Sharma--Mittal definition~\cite{SharmaMittal1975,NielsenNock2012}. For normalized states, $D^R_q$ is non-negative for $q\ge0$ and faithful for $q>0$~\cite{Renyi1961,vanErven2014}. At $\rho=\tI$, Petz and sandwiched forms coincide and obey data processing for $0\le q\le2$~\cite{Petz1986} and $q\ge\frac12$~\cite{FrankLieb2013}, respectively. For $q<0$, $D^R_q(\rho\Vert\sigma)=\frac{q}{1-q}D^R_{1-q}(\sigma\Vert\rho)\le0$~\cite{vanErven2014}. The associated Lindblad-completed Tsallis expression also has the reversed-divergence structure:
\begin{equation}
\TrF(\tN-\tI-\ln_q\tN)=\frac{q}{1-q}\Big[D^T_{1-q}(\tN\Vert\tI)+\TrF(\tI-\tN)\Big].
\label{eq:skew}
\end{equation}
This identity follows eigenvalue by eigenvalue from $D^T_{1-q}(\tN\Vert\tI)=\TrF(\tN-\tN^{1-q})/q$ and
\begin{align*}
x-1-\ln_qx&=\frac{x-qx+q-x^{1-q}}{1-q}\\
&=\frac{q}{1-q}\Big[\frac{x-x^{1-q}}{q}+1-x\Big].
\end{align*}
For $q<0$ the Tsallis action consists of its linear term minus a reversed-order divergence; its curvature is negative at the identity, as for $F(x)=x\ln x$. We therefore restrict the family to $q\ge0$ and real $r$, with the ghost-free branch $q=0$ on the boundary. Results quoted at $q<0$ (for example, $q=-1$ in Ref.~\cite{PaperII}) are formal continuations. These non-negativity statements apply to normalized states, $\Tr\rho=\Tr\sigma$. For the unnormalized $\tN$ used here, they apply if $\TrF\tN=\cN$; without that condition, the Lindblad-completed form $\TrF(\tN-\tI-\ln_q\tN)$ is non-negative for $q\ge0$, since $\ln_qx\le x-1$, and for $q>0$ vanishes only at $\tN=\tI$. The action itself remains sign-indefinite because of its linear term [Eq.~\eqref{eq:split}].

For $q=1$, all expressions below are understood as limits, with $\ln_1=\ln$ and $\bar Z_q^{\,s}\to e^{(r-1)\mathcal L/\cN}$. Define $\tau:=(1-q)/\cN$, so that $\bar Z_q=1-\tau\mathcal L^T_q$. The corresponding maps in Eq.~\eqref{eq:family} are
\begin{gather}
F=\ln_q,\qquad \Psi_{SM}(y)=\frac{\cN}{r-1}\big[(1-\tau y)^s-1\big],\label{eq:PsiSM}\\
\Psi_T(y)=y\ (r=q),\qquad \Psi_R(y)=-\frac1\tau\ln(1-\tau y)\ (r=1).\nonumber
\end{gather}
Differentiating gives
\begin{gather}
\Psi_{SM}'(y)=(1-\tau y)^{s-1}=\bar Z_q^{\,s-1},\qquad \Psi_{SM}''(0)=\frac{r-q}{\cN},\label{eq:Psider}\\
\Psi_{SM}''(y)=\frac{r-q}{\cN}(1-\tau y)^{s-2}.\nonumber
\end{gather}
Thus $\Psi_{SM}$ is convex if and only if $r\ge q$, and
\begin{equation}
\mathcal L^{SM}_{q,r}=\mathcal L^T_q+\frac{r-q}{2\cN}(\mathcal L^T_q)^2+O\big((\mathcal L^T_q)^3\big).
\label{eq:SMexp}
\end{equation}
Although each map $\Psi$ is monotone, a nonlinear choice changes the field equations: in general, varying $\int\sg\,\Psi(\mathcal L)$ is not equivalent to varying $\int\sg\,\mathcal L$.

\subsection{Special members and additivity}\label{sec:special}
\paragraph*{Tsallis at $q=0$: exact GR.} Because $\ln_0x=x-1$,
\begin{equation}
\mathcal L^T_0=-\TrF(\tN-\tI)=3\beta R-\alpha\TrF\tM\tg^{-1}
\label{eq:q0GR}
\end{equation}
exactly, to all orders in $\alpha$ and $\beta$. The Tsallis family therefore interpolates between GR ($q=0$) and GfE ($q=1$). At $q=0$, however, this functional is not a faithful divergence: $D^T_0(\tI\Vert\tN)=-\TrF(\tN-\tI)$ has no definite sign and vanishes whenever $\TrF\tN=\cN$. The ghost-free branch thus occurs where the divergence no longer distinguishes operators with equal trace.
\paragraph*{Tsallis at $q=\frac12$: Dirac--Born--Infeld.} For the warm-up rank-one spectrum $\{1^{(d-1)},1+\alpha w\}$, the Tsallis action is $\mathcal L^T_{1/2}=-2(\sqrt{1+\alpha|\nabla\phi|^2}-1)$, the DBI scalar associated with the induced metric $\bG$. GfE instead gives $-\ln(1+\alpha|\nabla\phi|^2)$.
\paragraph*{Additivity.} The functional $\mathcal L_F$ sums the three form-sector contributions $Z_{q,(n)}:=\TrM\mathbf N_{(n)}^{1-q}$, whereas nonlinear $\Psi$ acts on their total, $\bar Z_q=\cN^{-1}\sum_nZ_{q,(n)}$. Because R\'enyi entropy is additive for tensor products, not direct sums, R\'enyi and Sharma--Mittal (when $r\ne q$) couple the scalar, vector, and bivector sectors locally. The action remains additive over disjoint spacetime regions, since $\Psi$ acts pointwise and $\Psi(0)=0$; within each region, the coupling includes a cross term $\psi_2y_1y_2$ between contributions $y_1$ and $y_2$.

\section{Weak-field expansion and particle content}\label{sec:spectrum}
\subsection{Quadratic action}
With $\kappa_2=-F''(1)/2$ and $\psi_2=\Psi''(0)$, $-F(1+x)=-x+\kappa_2x^2+O(x^3)$ and $\Psi(y)=y+\frac{\psi_2}{2}y^2+O(y^3)$. In vacuum $Y=-\bigoplus_n\beta_n\mathbf X_{(n)}$, so
\begin{equation}
\mathcal L_{F,\Psi}=\beff R+\kappa_2\sum_n\beta_n^2\TrM\mathbf X_{(n)}^2+\frac{\psi_2}{2}\beff^2R^2+O(\mathcal R^3),
\label{eq:quadgen}
\end{equation}
with $\beff=\beta_0\gamma_{R0}+\beta_1\gamma_{R1}+\beta_2\gamma_{R2}$; only the scalar curvature components contribute to the trace (Appendix~\ref{app:weyl}).
Using the orthogonal decomposition of Appendix~\ref{app:weyl} and the Gauss--Bonnet density
\begin{align*}
\mathcal E&=R^2-4R_{\mu\nu}R^{\mu\nu}+R_{\mu\nu\rho\sigma}R^{\mu\nu\rho\sigma}\\
&=C^2-2S_{\mu\nu}S^{\mu\nu}+\tfrac16R^2,
\end{align*}
the quadratic part becomes
\begin{equation}
\mathcal L^{(2)}=\kappa_2\big[\mathcal W C^2+\mathcal Q R^2-\tfrac12\Sigma\,\mathcal E\big]+\frac{\psi_2}{2}\beff^2R^2,
\label{eq:quad}
\end{equation}
where
\begin{align}
\mathcal W&=\beta_2^2\gamma_W^2+\tfrac12\beta_1^2\gamma_{S1}^2+\beta_2^2\gamma_{S2}^2\ \ \ge0,\label{eq:W}\\
\mathcal Q&=\beta_0^2\gamma_{R0}^2+\beta_1^2\Big(\frac{\gamma_{R1}^2}4+\frac{\gamma_{S1}^2}{12}\Big)+\beta_2^2\,\frac{\gamma_{R2}^2+\gamma_{S2}^2}{6},\label{eq:Q}\\
\Sigma&=\beta_1^2\gamma_{S1}^2+2\beta_2^2\gamma_{S2}^2 .
\end{align}
In $d=4$, $\int\sg\,\mathcal E$ is topological (Chern--Gauss--Bonnet) and $\Sigma$ drops out. The GfE blocks (all $\gamma=1$, $\beta_n=\beta$) have $\beff=3\beta$, $\mathcal W=\frac52\beta^2$ and $\mathcal Q=\frac53\beta^2$; equivalently, by Eq.~\eqref{eq:TrX2},
\begin{equation}
\mathcal L_{\rm GfE}=3\beta R+\frac{\beta^2}{2}\big(\mathcal E+5R_{\mu\nu}R^{\mu\nu}\big)+O(\beta^3\mathcal R^3).
\label{eq:quadGfE}
\end{equation}
For all three entropies $\kappa_2=q/2$, while $\psi_2=0$ (Tsallis), $(1-q)/\cN$ (R\'enyi) and $(r-q)/\cN$ (Sharma--Mittal). Thus, the additional $R^2$ term proportional to $\psi_2$ below is absent for Tsallis:
\begin{align}
\mathcal L^{SM}_{\rm vac}={}&3\beta R+\frac{q\beta^2}{2}(\mathcal E+5R_{\mu\nu}R^{\mu\nu})\nonumber\\
&+\frac{9(r-q)\beta^2}{2\cN}R^2+O(\mathcal R^3).
\label{eq:quadSM}
\end{align}
\paragraph*{Universal Einstein limit.} Every member has the same linear term, $\beff R-\sum_n\alpha_n\TrM\bM_{(n)}g^{-1}_{(n)}$. Thus Newton's constant and the leading matter operator, including its $\xi R$ coupling, are independent of $(q,r)$; differences arise at higher order.
\paragraph*{Matter at second order.} Matter introduces a curvature--matter term $-2\kappa_2\alpha\beta\,\TrF(\tM\tg^{-1}\tX)$ and quartic self-interactions $\kappa_2\alpha^2\TrF(\tM\tg^{-1})^2$. For $\psi_2\ne0$, there is also a term $3\psi_2\alpha\beta R\,\mathcal L_m$, coupling curvature to the full matter Lagrangian. These terms are suppressed by $\beta\mathcal R$ or $\rho/\rho_P$, with $\rho_P:=M_P^4$. Their implications for free fall and light propagation are studied in Ref.~\cite{PaperII}.

\subsection{Linear spectrum}
Since cubic curvature terms do not contribute to the quadratic action about Minkowski space, Eq.~\eqref{eq:quad} fixes the linear spectrum. Write the vacuum Lagrangian as $\beff[R+a_CC^2+a_RR^2]$, up to the Gauss--Bonnet term. Appendix~\ref{app:lin} shows that transverse-traceless (TT) perturbations and the conformal mode obey
\begin{equation}
\Box(1+2a_C\Box)h^{TT}_{\mu\nu}=0,\qquad \Box(1-6a_R\Box)\vartheta=0 ,
\label{eq:linops}
\end{equation}
where $g=\eta+h$, with $\eta$ the Minkowski metric and $h^{TT}_{\mu\nu}$ the transverse-traceless part of $h_{\mu\nu}$, while $g=e^{2\vartheta}\eta$ defines the conformal mode.
The TT propagator decomposes as
\begin{equation}
\frac{1}{\Box(1+2a_C\Box)}=\frac1\Box-\frac{1}{\Box+1/(2a_C)}
\end{equation}
and contains a massless graviton and a massive spin-2 pole with opposite residue; the latter is a ghost~\cite{Stelle1977,Stelle1978}. Its squared mass is $m_2^2=-1/(2a_C)$ in the convention $(\Box-m_2^2)h=0$. A ghost has the wrong-sign residue, while a tachyon has negative squared mass. The scalar mass is $m_0^2=1/(6a_R)$. Its residue is opposite to that of the constrained, wrong-sign conformal mode, so it is not a ghost and is healthy when $m_0^2>0$. Using $a_C=\kappa_2\mathcal W/\beff$ and $a_R=(\kappa_2\mathcal Q+\frac12\psi_2\beff^2)/\beff$, we find
\begin{equation}
m_2^2=-\frac{\beff}{2\kappa_2\mathcal W},\qquad m_0^2=\frac{\beff}{6\kappa_2\mathcal Q+3\psi_2\beff^2}.
\label{eq:masses}
\end{equation}
For GfE this agrees with the standard quadratic-gravity masses~\cite{Stelle1978,DeserTekin2003b,Bueno2017}.

\paragraph{GfE.} With $\beta'=1/48\pi$,
\begin{equation}
m_0^2=\frac{3}{5\beta}\simeq90\,M_P^2,\qquad m_2^2=-\frac{6}{5\beta}\simeq-181\,M_P^2.
\label{eq:massesGfE}
\end{equation}
Around flat space GfE therefore propagates the graviton, a healthy scalaron as in $R+R^2$ gravity~\cite{Starobinsky1980}, and a Planck-scale spin-2 ghost with negative squared mass. \rev{This spectrum was first obtained by Pereira~\cite{Pereira2026a}, who also showed} that every Ricci-flat metric solves the vacuum equations through quadratic order in curvature\rev{; our derivation agrees with his}. \rev{This bears on the Ostrogradsky question raised in Ref.~\cite{GfE}: although the G-field equations are of second order, the ghost remains in the spectrum~\cite{Woodard2015}. Its nonlinear fate requires a Hamiltonian analysis, which would also be the starting point for the canonical quantization envisaged in Ref.~\cite{GfE}.} The masses depend on couplings only through $\beff$ and squared combinations of $\beta_n\gamma$, and are invariant under reversal of the curvature convention provided the Einstein--Hilbert term remains positive; this reversal also changes the sign of $\beta$ and leaves $\tN$ and $\mathcal L$ unchanged.

\paragraph*{Status of the ghost.} Planck-scale ghosts also occur in quadratic gravity~\cite{Stelle1977,Stelle1978,Salvio2018}, though their interpretation remains debated~\cite{LeeWick1969,Anselmi2017,DonoghueMenezes2019}. If GfE is an effective field theory below $|m_2|$, perturbative order reduction removes the extra poles~\cite{PaperII}. If the quadratic action is instead treated as exact, the tachyon destabilizes Minkowski space on a timescale $|m_2|^{-1}\sim\lP$.

\paragraph*{Sharma--Mittal.} Inserting $\kappa_2=q/2$ and $\psi_2=(r-q)/\cN$ from Eq.~\eqref{eq:quadSM} into Eq.~\eqref{eq:masses} gives, in $d=4$ (Appendix~\ref{app:lin}),
\begin{equation}
m_2^2=-\frac{6}{5q\beta},\qquad m_0^2\overset{\cN=11}{=}\frac{33}{\beta(27r+28q)}.
\label{eq:massesSM}
\end{equation}

\begin{table*}[t]
\caption{Members of the \rev{GfGE} family~\eqref{eq:family} with the GfE curvature blocks, except in the last row. Here $\kappa_2=-F''(1)/2$ and $\psi_2=\Psi''(0)$; masses are for $\cN=11$ ($d=4$); ``$\Lambda_{q,r}\ge0$'' refers to the G-field potential~\eqref{eq:Lqr} (Sec.~\ref{sec:potentials}); the last row is discussed in Sec.~\ref{sec:ghostfree}.}
\label{tab:family}
\begin{ruledtabular}
\begin{tabular}{lccccccl}
Theory & $F$ & $\Psi$ & $\kappa_2$ & $\psi_2$ & $m_2^2$ & $m_0^2$ & Remarks\\
\hline
GfE~\cite{GfE} & $\ln$ & id & $\frac12$ & $0$ & $-\frac{6}{5\beta}$ & $\frac{3}{5\beta}$ & $\Lambda_{q,r}\ge0$\\
Tsallis & $\ln_q$ & id & $\frac q2$ & $0$ & $-\frac{6}{5q\beta}$ & $\frac{3}{5q\beta}$ & $\mathrm{sign}\,\Lambda_{q,r}=\mathrm{sign}\,q$; $q=0$: exact GR\\
R\'enyi & $\ln_q$ & $\Psi_R$ & $\frac q2$ & $\frac{1-q}{\cN}$ & $-\frac{6}{5q\beta}$ & $\frac{33}{\beta(27+28q)}$ & $\Lambda_{q,r}\ge0$ for $q>0$; $q=0$: ghost-free $f(R,\mathcal L_m)$\\
Sharma--Mittal & $\ln_q$ & $\Psi_{SM}$ & $\frac q2$ & $\frac{r-q}{\cN}$ & $-\frac{6}{5q\beta}$ & $\frac{33}{\beta(27r+28q)}$ & $\Lambda_{q,r}\ge0$ for $q,r>0$; $q=0$, $r>0$: ghost-free\\
Trace-projected GfE & $\ln$ & id & $\frac12$ & $0$ & none & $\frac{12}{17\beta}$ & $\Lambda_{q,r}\ge0$; vacuum $f(R)$, Eq.~\eqref{eq:ftrace}\\
\end{tabular}
\end{ruledtabular}
\end{table*}

The masses for the Tsallis and R\'enyi limits are listed in Table~\ref{tab:family}. For every $r$, $m_2^2$ scales as $1/q$, so $q\to0^+$ decouples the massive spin-2 mode, which is absent at $q=0$. \rev{Along the Tsallis line $r=q$ the scalar mass scales as $1/q$ as well and also decouples, whereas at fixed $r>0$ it tends to $m_0^2=\cN/(9r\beta)$.}

\subsection{No-go result for massive spin-2 modes}\label{sec:nogo}
For the family in Eqs.~\eqref{eq:family}--\eqref{eq:blocks}, with $\beff\ne0$, a massive spin-2 mode occurs around Minkowski space if and only if $\kappa_2\mathcal W\ne0$. Equivalently, $F''(1)\ne0$ and at least one of $\beta_1\gamma_{S1}$, $\beta_2\gamma_W$, or $\beta_2\gamma_{S2}$ is nonzero. Whenever present, the mode is a ghost and is tachyonic precisely when $\kappa_2\beff>0$. Indeed, Eq.~\eqref{eq:masses} gives a pole when $a_C=\kappa_2\mathcal W/\beff\ne0$; $\mathcal W$ is a positively weighted sum of squares and vanishes only when all three curvature couplings vanish. The propagator's partial fractions fix the residue sign, while $\mathrm{sign}\,m_2^2=-\mathrm{sign}(\kappa_2\beff)$. Since $\Psi$ contributes only to $R^2$ at this order, it cannot change this conclusion.

For a common generator $F$ across sectors, absence of the massive spin-2 ghost therefore has two branches: (a) a \emph{trace-projected} branch, in which curvature enters only through $R$, and (b) a \emph{flat-$F$} branch with $F''(1)=0$. For $F=\ln_q$, the latter is $q=0$ and gives an exact $f(R,\mathcal L_m)$ theory.
If a general $F$ has $F'''(1)\ne0$, cubic curvature terms remain and its effective second derivative need not vanish away from $\tN=\tI$; ghost freedom is then established only around Minkowski space. It holds on arbitrary backgrounds if $F''$ vanishes identically, as for $\ln_0$. On this $q=0$ branch, Tsallis is general relativity, while R\'enyi and Sharma--Mittal have healthy scalar masses $\cN/(9\beta)$ and $\cN/(9r\beta)$ for $r>0$.

\paragraph*{Sector-dependent generating functions.} The no-go result assumes a common $F$ in all sectors. If the generating functions are sector dependent, contributions to $\mathcal W$ can cancel when the values of $F_n''(1)$ have opposite signs. Such a cancellation requires mixing entropies with opposite concavity across form degrees.

\begin{figure}[t]
\includegraphics[width=\columnwidth]{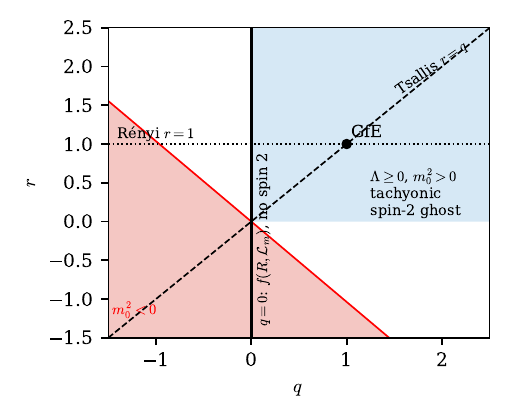}
\caption{Sharma--Mittal GfE spectrum in the $(q,r)$ plane. In blue, the scalar is healthy and the potential is non-negative, but a tachyonic spin-2 ghost is present. Below the red line $27r+28q=0$, the scalar is tachyonic. At $q=0$ the massive spin-2 mode is absent. The dashed diagonal is Tsallis, the dotted horizontal line is R\'enyi, and the dot is GfE. The region $q<0$ is a formal continuation beyond the non-negative-divergence domain; for $q>0,r<0$, potential non-negativity is not guaranteed.}
\label{fig:phase}
\end{figure}

\paragraph*{Sign of the scalar mass.} For concave $F$ ($\kappa_2>0$) and $\psi_2\ge0$, the scalar is healthy. If $\kappa_2<0$, the spin-2 ghost is non-tachyonic, but the scalar is tachyonic unless $\psi_2$ is sufficiently large to compensate. The scalar is tachyonic for $q<0$ in Tsallis, for $q<-\frac{27}{28}$ in R\'enyi, and when $27r+28q<0$ in Sharma--Mittal [Fig.~\ref{fig:phase}]. Within the adopted domain $q\ge0$, only the Sharma--Mittal case occurs: it requires $r<-28q/27$, where the negative $\psi_2$ contribution dominates. The other tachyonic regions are formal continuations.

\section{\texorpdfstring{G-field formulation of \rev{GfGE}}{G-field formulation of GfGE}}\label{sec:Gfield}
\subsection{Auxiliary fields and Bregman potential}
Generalizing Eq.~\eqref{eq:Ltilde}, let $L(\tTh):=\Psi(-\TrF F(\tTh))$ and
\begin{equation}
\tilde{\mathcal L}=L(\tTh)-\TrF\tGc(\tN-\tTh).
\label{eq:Ltildegen}
\end{equation}
The auxiliary equations are $\tTh=\tN$ and
\begin{equation}
\tGc=-\frac{\partial L}{\partial\tTh}=\Omega\,F'(\tTh),\qquad \Omega:=\Psi'\big(-\TrF F(\tTh)\big).
\label{eq:Ggen}
\end{equation}
On shell $\tGc=\Omega F'(\tN)$, and since both equations are algebraic, the equivalence established in Sec.~\ref{sec:equiv} continues to hold: $\tilde{\mathcal L}$ and $\mathcal L_{F,\Psi}$ have the same classical solutions. Expanding $\tN$ using Eq.~\eqref{eq:Ngen} and solving Eq.~\eqref{eq:Ggen} for $\tTh(\tGc)$ gives
\begin{gather}
\tilde{\mathcal L}=\sum_n\beta_n\TrM\bm{\mathcal G}_{(n)}\mathbf X_{(n)}-V(\tGc)\nonumber\\
-\sum_n\alpha_n\TrM\bm{\mathcal G}_{(n)}\bM_{(n)}g_{(n)}^{-1},\label{eq:dressedgen}\\
V(\tGc)=\TrF\tGc-\TrF\tGc\tTh-L(\tTh).\nonumber
\end{gather}
\paragraph*{Bregman potential.} Using $L(\tI)=0$ and $\partial L(\tTh)=-\tGc$, the potential is $V=L(\tI)-L(\tTh)-\TrF[\partial L(\tTh)(\tI-\tTh)]$, the Bregman divergence of $L$ between $\tI$ and $\tTh$. Since $\tI$ commutes with $\tTh$, it depends only on the eigenvalues of $\tTh$. It is non-negative whenever $L(\tI)=0$ lies above the tangent plane of $L$ at each point in its domain; convexity of $L$ in the eigenvalues is a sufficient condition. For $\Psi=\mathrm{id}$, $V=\TrF[F(\tTh)+F'(\tTh)(\tI-\tTh)]$, and a sufficient condition is that $F$ be concave, since a concave function lies below its tangent line, $F(1)\le F(\lambda)+F'(\lambda)(1-\lambda)$. Then
\begin{equation}
V=\sum_\lambda\big[F(\lambda)+F'(\lambda)(1-\lambda)\big]\ge0 ,
\label{eq:bregmanF}
\end{equation}
where $\lambda$ runs over the spectrum of $\tTh$. For GfE this is $V=\TrF(\tGc-\tI-\ln\tGc)=2\beta\LG$.
On shell, the dressed metric is $\tg_{\mathcal G}=\tGc^{-1}\tg=\Omega^{-1}F'(\tN)^{-1}\tg$. For $F=\ln_q$ $F'(\tN)^{-1}=\tN^q$, so
\begin{equation}
\tg_{\mathcal G}=\Omega^{-1}\,\tN^q\tg=\Omega^{-1}\,\tg\,\#_q\,\tG,
\label{eq:geomean}
\end{equation}
where, for $0\le q\le1$, $\#_q$ is the Kubo--Ando weighted geometric mean~\cite{KuboAndo1980}, formally $\tg^{1/2}(\tg^{-1/2}\tG\tg^{-1/2})^q\tg^{1/2}$. Since $\tg^{1/2}$ does not exist for Lorentzian $\tg$, we define $\tg\,\#_q\tG:=\tN^q\tg$, which coincides with the Kubo--Ando mean whenever both metrics are positive definite. For $0\le q\le1$, matter thus propagates in a metric that interpolates geometrically between $\tg$ ($q=0$) and $\tG$ ($q=1$), while for $q>1$ the same expression extrapolates beyond $\tG$. The factor $\Omega$ is common to all sectors and multiplies the curvature and matter terms alike, like a Jordan-frame conformal factor; being a field, it also contributes through its gradients to the derivative terms $\mathcal D_{\mu\nu}$ of the field equations (Sec.~\ref{sec:fieldeqs}).

\subsection{Tsallis, R\'enyi and Sharma--Mittal potentials}\label{sec:potentials}
For $F=\ln_q$ and $\Psi=\Psi_{SM}$, Eq.~\eqref{eq:Ggen} gives
\begin{equation}
\tGc=\bar Z_q^{\,s-1}\,\tTh^{-q},
\label{eq:escort}
\end{equation}
where $\bar Z_q=\av{\tTh^{1-q}}$ is evaluated on $\tTh$. For Tsallis ($s=1$) this is $\tGc=\tTh^{-q}$, and for R\'enyi ($s=0$) it is $\tGc=\tTh^{-q}/\av{\tTh^{1-q}}$, so that $\tGc\tTh=\tTh^{1-q}/\av{\tTh^{1-q}}$ is the operator analogue of an \emph{escort distribution} $p^{(q)}_i=p_i^q/\sum_jp_j^q$ of order $1-q$~\cite{BeckSchlogl,TsallisBook,Naudts2011}. The R\'enyi G-field satisfies $\TrF\tGc\tTh=\cN$ (the same trace is $\cN\bar Z_q$ for Tsallis and $\cN\bar Z_q^{\,s}$ for Sharma--Mittal), so that on shell $\TrF(\tg_{\mathcal G}^{-1}\tG)=\cN$: the R\'enyi G-field renormalizes the ``renormalizable density matrix'' $\tG$ to unit average eigenvalue relative to the dressed metric. For Tsallis ($\Omega=1$) the dressed metric~\eqref{eq:geomean} is $\tg\,\#_q\,\tG$ itself, whereas for R\'enyi and Sharma--Mittal it is this geometric mean multiplied by the conformal factor $\Omega^{-1}=\bar Z_q$ and $\bar Z_q^{\,1-s}$, respectively.

\paragraph*{Power-mean form of the cosmological term.} Let $\mathsf M_p(\tGc):=\av{\tGc^p}^{1/p}$ be the power means of the spectrum of $\tGc$, and set $\varepsilon=1-1/q$. For Sharma--Mittal GfE with $q\neq0$ and $r\neq0$, the potential takes the form (Appendix~\ref{app:lambda})
\begin{equation}
2\beta\Lambda_{q,r}:=V=\cN\Big[\mathsf M_1(\tGc)-1-\ln_{1/r}\mathsf M_\varepsilon(\tGc)\Big].
\label{eq:Lqr}
\end{equation}
Its special cases are $V=\TrF(\tGc-\tI-\ln_{1/q}\tGc)$ for Tsallis, $V=\cN[\mathsf M_1-1-\ln\mathsf M_\varepsilon]$ for R\'enyi and $V=\TrF(\tGc-\tI-\ln\tGc)$ for GfE.
If $q>0$ and $r>0$ (for Tsallis and R\'enyi, $q>0$), then $\Lambda_{q,r}\ge0$, with equality iff $\tGc=\tI$. For $q>0$ we have $\varepsilon<1$, and the power-mean (Jensen) inequality gives $\mathsf M_\varepsilon\le\mathsf M_1$; since $\ln_{1/r}$ is increasing, and concave for $r>0$, it follows that $\ln_{1/r}\mathsf M_\varepsilon\le\ln_{1/r}\mathsf M_1\le\mathsf M_1-1$. Equality requires all eigenvalues to equal a common $c$ with $c-1=\ln_{1/r}c$, i.e.\ $c=1$.
At $q=0$ the map $\tTh\mapsto\tGc$ is not invertible: $\tGc=\chi\tI$ with $\chi=\bar Z_0^{\,-r}$ ($\chi=1$ for Tsallis), and $V=\cN(\chi-1-\ln_{1/r}\chi)$ on this ray (Sec.~\ref{sec:ghostfree}). At $r=0$ the G-field is constrained by $\mathsf M_\varepsilon(\tGc)=1$, and $V=\cN[\mathsf M_1(\tGc)-1]$.
\paragraph*{Interpretation.} Two dual indices appear, $r\to1/r$ in the deformed logarithm and $q\to1-1/q$ in the power mean. In the Tsallis case the potential is the Lindblad-completed Tsallis divergence $D^T_{1/q}(\tI\Vert\tGc)+\TrF(\tGc-\tI)$, whose sign is that of $q$. As for GfE (Sec.~\ref{sec:equiv}), $\Lambda_{q,r}$ does not act as a cosmological constant on shell. In vacuum $\tGc=\tI+\beta(2\kappa_2\tX+3\psi_2R\,\tI)+O(\beta^2)$, so $\beta\TrF\tGc\tX-V=3\beta R+O(\beta^2)$, and $\Lambda_{q,r}$ contributes only to the curvature-squared terms.

\subsection{Field equations}\label{sec:fieldeqs}
The generalized entropies affect the field equations only through the algebraic G-field equation~\eqref{eq:Ggen} and the potential $V$, so the equations have the same form for Tsallis, R\'enyi and Sharma--Mittal. Varying Eq.~\eqref{eq:dressedgen} gives three sets of equations.
\paragraph*{Matter.} Since $d$ and $\delta$ are adjoint with respect to $\int\sg\langle\cdot,\cdot\rangle$, varying with respect to $\bra\Phi$ gives, for $\alpha_n=\alpha$,
\begin{equation}
\mathbb D\,\tg_{\mathcal G}^{-1}\mathbb D\ket\Phi+\tg_{\mathcal G}^{-1}(m^2+\xi R)\ket\Phi=0,\qquad \tg_{\mathcal G}^{-1}=\tg^{-1}\tGc .
\label{eq:mattereq}
\end{equation}
For $\tGc=\tI$ and $\omega=\zeta=0$ this is $(-\Box+m^2+\xi R)\phi=0$.
\paragraph*{Metric.} The G-field equation is the algebraic relation $\tGc=\Omega F'(\tN)$. For the metric equation we define, for $\alpha_n=\alpha$,
\begin{gather*}
\RG:=\sum_n\frac{\beta_n}{\beta}\TrM\bm{\mathcal G}_{(n)}\mathbf X_{(n)},\qquad \kappa_m:=\frac{\alpha}{\beta},\\
\mathcal S_m:=-\int\sg\,\MG\,\dr,\qquad \mathcal T_{\mu\nu}:=-\frac1{\sg}\frac{\delta\mathcal S_m}{\delta g^{\mu\nu}},
\end{gather*}
where $\beta$ is a reference value of the couplings $\beta_n$ ($\beta_n=\beta$ for the GfE blocks), $\RG$ generalizes the dressed curvature scalar of Eq.~\eqref{eq:dressed} to sector couplings and $\mathcal T_{\mu\nu}$ is one half of the standard stress tensor $T_{\mu\nu}=-\frac2{\sg}\frac{\delta\mathcal S_m}{\delta g^{\mu\nu}}$. Appendix~\ref{app:var} then gives
\begin{equation}
R^{\mathcal G}_{(\mu\nu)}-\tfrac12g_{\mu\nu}\Big(\RG-\frac{V}{\beta}\Big)+\mathcal D_{(\mu\nu)}=\kappa_m\,\mathcal T_{(\mu\nu)} .
\label{eq:modEin}
\end{equation}
Here $R^{\mathcal G}_{\mu\nu}$ is the dressed Ricci tensor and $\mathcal D_{\mu\nu}$ collects second derivatives of $\tGc$; both are defined in Eq.~\eqref{eq:RGD} of Appendix~\ref{app:var}, where we also show that $\beta\tGc$ is the tensor $\partial\mathcal L/\partial R_{\mu\nu\rho\sigma}$. Because $V$ depends only on the mixed components of $\tGc$, it enters only through $\sg$. The system is of second order in $g$, $\Phi$ and $\tGc$, and for $\tGc\to\tI$ it reduces to $3G_{\mu\nu}= \kappa_m\mathcal T_{\mu\nu}$ for every member of the family, i.e.\ to $G_{\mu\nu}=8\pi GT_{\mu\nu}$ by Eq.~\eqref{eq:norm}. Diffeomorphism invariance of the G-field action $\tilde{\mathcal S}$ guarantees the consistency of the full system (generalized Bianchi identity). Since matter couples to $\tGc$, however, $\mathcal T_{\mu\nu}$ is in general not separately conserved: $\nabla^\mu\mathcal T_{\mu\nu}$ is proportional to gradients of $\tGc$, describing energy--momentum exchange between matter and the G-field, as in other theories with non-minimal matter couplings~\cite{HarkoLobo2010}.

\paragraph*{On the physical status of the G-field.} Reference~\cite{GfE} advocates regarding \rev{the multipliers} $\tGc$ as physical fields that add new dimensions to the phase space of the equations of motion. The classical equivalence of Sec.~\ref{sec:equiv} makes this view precise. Eliminating $\tGc$ and $\tTh$ through the algebraic equations~\eqref{eq:aux} gives a metric equation with higher derivatives of $g$, through $\mathcal D_{\mu\nu}$ [Eq.~\eqref{eq:modEin}] and $\tGc^{-1}=\tN$, whereas in the G-field formulation the equations are of second order in $g$ and $\tGc$, as for $f(R)$ gravity in its Brans--Dicke form~\cite{SotiriouFaraoni2010}\rev{, an analogy drawn in Ref.~\cite{GfE}}. The extra initial data of the higher-derivative metric formulation therefore correspond to data of the G-field. At the linearized level the counting is explicit: around Minkowski space the theory propagates $2+1+5=8$ degrees of freedom (Sec.~\ref{sec:spectrum}), as in quadratic gravity~\cite{Stelle1978}, and the six beyond the graviton are fluctuations of the G-field. The same counting holds for the Tsallis, R\'enyi and Sharma--Mittal deformations with $q\neq0$, whereas the ghost-free branches of Sec.~\ref{sec:ghostfree} propagate $2+1$ (only the graviton for Tsallis at $q=0$, which is GR). In this sense the G-field is physical, but the number and nature of its modes are fixed by the original action; changing them would require independent kinetic terms for $\tGc$, i.e.\ a different theory.

\subsection{Horizon entropy}\label{sec:horizon}
The Wald-tensor identity of Sec.~\ref{sec:equiv} holds for every member of Eq.~\eqref{eq:family} with GfE curvature blocks, because Eq.~\eqref{eq:PG} follows from the algebraic G-field equation~\eqref{eq:Ggen}. For the general blocks~\eqref{eq:blocks}, the contributions carry the corresponding $\beta_n/\beta$ and $\gamma$ weights; the entropy choice determines the relation between $\tGc$ and curvature through Eq.~\eqref{eq:Ggen}. This identity also explains the Padmanabhan form of Eq.~\eqref{eq:modEin}~\cite{Padmanabhan2013}. Equivalently, curvature enters the auxiliary action linearly with coefficient $\beta\tGc$; eliminating the algebraic auxiliary fields reproduces the original on-shell Wald entropy~\cite{JacobsonKangMyers1994}. Because the binormal has the algebraic symmetries of the Riemann tensor, the projector $\mathbb P$ drops out of the entropy contraction. In $d=4$, with $\beta'=1/48\pi$, Eq.~\eqref{eq:PG} then gives
\begin{gather}
S_W=\frac{A}{4G}\,\big\langle\mathcal G_\perp\big\rangle_{\mathcal H},\nonumber\\
\mathcal G_\perp:=\tfrac16\big[2\bm{\mathcal G}_{(0)}+\Tr_\perp\bm{\mathcal G}_{(1)}+2\bm{\mathcal G}_{(2)}(n)\big],
\label{eq:SWperp}
\end{gather}
where $A$ is the area of the bifurcation surface $\mathcal H$, $\av{\cdot}_{\mathcal H}$ the average over it, $\Tr_\perp$ the trace of the 1-form block over the two-plane normal to the horizon, and $\bm{\mathcal G}_{(2)}(n):=-\frac12n_{\mu\nu}\bm{\mathcal G}_{(2)}{}^{\mu\nu\rho\sigma}n_{\rho\sigma}$ the normalized expectation value of the 2-form block in the binormal bivector. For $\tGc=\tI$ this is the Bekenstein--Hawking entropy $A/4G$~\cite{Bekenstein1973,Hawking1975}. In general, the departure of the horizon entropy from the area law measures the departure of the G-field from the identity, i.e.\ the distinguishability of the induced and spacetime metrics, in the directions normal to the horizon.

As a check, consider a Schwarzschild black hole. Since $\mathbf X_{(0)}=\mathbf X_{(1)}=0$ on Ricci-flat metrics, to first order in $\beta$ only the 2-form block of the G-field is non-trivial. The $t$--$r$ bivector is an eigenvector of $\tX$ with eigenvalue $\lambda=4GM/r^3$ at areal radius $r$ (not to be confused with the Sharma--Mittal index)~\cite{BianconiBH2025,PaperII}, and Eq.~\eqref{eq:escort} gives $\bm{\mathcal G}_{(2)}(n)=(1-\beta\lambda)^{-q}[1+O(\beta^2)]$ for all three entropies. At the horizon, $r=2GM$, Eq.~\eqref{eq:SWperp} then yields $S_W=A/4G+4\pi\kappa_2\beta'/3$ to first order. This is the Jacobson--Myers entropy~\cite{JacobsonMyers1993} of the Gauss--Bonnet term $\kappa_2\beta^2\mathcal E$ contained in $\kappa_2\beta^2\TrF\tX^2$, in agreement with the full calculation of Ref.~\cite{PaperII}. \rev{This Wald entropy is distinct from the area law of Ref.~\cite{BianconiBH2025}, obtained by integrating the quantum relative entropy over the black-hole interior during a finite time interval; the two notions are complementary.}

More generally, since $F$ and $\Psi$ are analytic at the identity with $F'(1)=\Psi'(0)=1$, the G-field is $\tGc=\tI+O(\beta\mathcal R)$, and Eq.~\eqref{eq:SWperp} gives $S_W=(A/4G)[1+O(\beta\mathcal R_{\mathcal H})]$, with $\mathcal R_{\mathcal H}$ the curvature at the horizon. For horizons with a single curvature scale, $\mathcal R_{\mathcal H}\propto1/A$, as for Schwarzschild and de Sitter, this is the area law followed by a constant and inverse powers of $A$, for every value of the indices. The entropic deformation of the action therefore does not produce the power-law horizon entropies $S\propto A^\delta$ of Tsallis and Cirto~\cite{TsallisCirto2013} or $S\propto A^{1+\Delta/2}$ of Barrow~\cite{Barrow2020}; the explicit coefficients and the consequences for horizon thermodynamics are given in Ref.~\cite{PaperII}.

\section{Ghost-free branches}\label{sec:ghostfree}
\subsection{Trace-projected GfE}
By the no-go result of Sec.~\ref{sec:nogo}, the first ghost-free branch projects the curvature blocks onto their scalar traces, $\mathbf X_{(n)}=\gamma_{Rn}\frac{R}{d_n}\mathbf 1_{d_n}$, with $d_n=\binom4n$. Choosing $\gamma_{Rn}=1$ and $\beta_n=\beta$ preserves the GfE Einstein limit. It replaces
\begin{equation}
R_{\mu\nu}\to\tfrac R4g_{\mu\nu},\qquad R_{\mu\nu\rho\sigma}\to\tfrac{R}{12}(g_{\mu\rho}g_{\nu\sigma}-g_{\mu\sigma}g_{\nu\rho})
\end{equation}
in $\tR$. In vacuum, with $\Psi=\mathrm{id}$, the theory is the $f(R)$ theory~\cite{SotiriouFaraoni2010,DeFeliceTsujikawa2010}
\begin{align}
f(R)&=-\sum_nd_nF\Big(1-\frac{\beta R}{d_n}\Big)\label{eq:ftrace}\\
&\overset{F=\ln}{=}-\ln(1-\beta R)-4\ln\big(1-\tfrac{\beta R}4\big)-6\ln\big(1-\tfrac{\beta R}6\big).\nonumber
\end{align}
For increasing, concave $F$, $f'=\beta\sum_nF'(\cdot)>0$ and $f''=-\beta^2\sum_nF''(\cdot)/d_n>0$. The theory is then free of the scalar ghost and the Dolgov--Kawasaki instability~\cite{DolgovKawasaki2003}. For $\mathcal W=0$ and $\mathcal Q=\frac{17}{12}\beta^2$, Eq.~\eqref{eq:masses} gives the scalaron mass
\begin{equation}
m_0^2=\frac{12}{17\beta}\qquad(F=\ln).
\end{equation}
With matter the theory becomes a nonminimally coupled $f(R,\Phi)$ model because the matter blocks are not proportional to the identity. In vacuum, each $\bm{\mathcal G}_{(n)}=\chi_n\mathbf 1$, and the potential $\sum_nd_n(\chi_n-1-\ln\chi_n)$ retains its relative-entropy form.

\subsection{The line \texorpdfstring{$q=0$}{q=0}: exact \texorpdfstring{$f(R,\mathcal L_m)$}{f(R,Lm)} theories}
At $q=0$, the action depends only on $\bar Z_0=\av{\tN}=1-y/\cN$, where $y:=3\beta R+\alpha\mathcal L_m=-\TrF(\tN-\tI)$ and $\mathcal L_m$ is defined in Sec.~\ref{sec:weak}. Since each GfE curvature block has trace $R$ (Sec.~\ref{sec:topo}), the Ricci and Riemann tensors cancel from the action exactly. The result is
\begin{equation}
\mathcal L^{SM}_{0,r}=f^{(r)}(y)=\frac{\cN}{r-1}\Big[\Big(1-\frac y\cN\Big)^{1-r}-1\Big].
\label{eq:fr}
\end{equation}
This defines an $f(R,\mathcal L_m)$ theory~\cite{HarkoLobo2010}. The limits $r=0$, $r=1$, and $r=2$ give, respectively, $f=y$ (general relativity, corresponding to Tsallis at $q=0$), $f=-\cN\ln(1-y/\cN)$ (R\'enyi), and $f=y/(1-y/\cN)$ (the $r=2$ model). Differentiating gives
\begin{gather}
f^{(r)\prime}(y)=\Big(1-\frac y\cN\Big)^{-r}>0,\nonumber\\
f^{(r)\prime\prime}(y)=\frac r\cN\Big(1-\frac y\cN\Big)^{-r-1}.
\end{gather}
For every $r>0$, $f'>0$ and $f''>0$, so the theory is ghost-free. The scalaron mass is
\begin{equation}
m_0^2=\left.\frac{\partial_R f}{3\partial_R^2 f}\right|_{R=0}=\frac{f^{(r)\prime}(0)}{9\beta f^{(r)\prime\prime}(0)}=\frac{\cN}{9r\beta}.
\end{equation}
For $r=1$, the R\'enyi function has the auxiliary representation
\begin{equation}
f_1(y)=\mathrm{stat}_\chi\big[\chi y-\cN(\chi-1-\ln\chi)\big],
\end{equation}
where $\mathrm{stat}_\chi$ denotes evaluation at the stationary point. Varying with respect to $\chi$ gives $y=\cN(1-1/\chi)$; substitution yields $\chi y-\cN(\chi-1-\ln\chi)=\cN\ln\chi=-\cN\ln(1-y/\cN)$. Since $\TrF(\chi\tI-\tI-\ln\chi\tI)=\cN(\chi-1-\ln\chi)$, this is precisely the GfE auxiliary-field Lagrangian~\eqref{eq:dressed} restricted to $\tGc=\chi\tI$.

The spin-2 ghost of GfE is carried by the traceless Ricci and Weyl components of the operator-valued G-field. At $q=0$, only its trace remains: a Brans--Dicke-like scalar~\cite{SotiriouFaraoni2010} with $\omega_{BD}=0$ and potential $\cN(\chi-1-\ln\chi)=D(\tI\Vert\chi\tI)$. The entropic structure therefore survives while the spin-2 ghost is removed.

\paragraph{Field equations on the line $q=0$.} The G-field is now a scalar, $\tGc=f^{(r)\prime}(y)\,\tI$, and the equations of Sec.~\ref{sec:fieldeqs} reduce to those of the $f(R,\mathcal L_m)$ theories~\eqref{eq:fr}~\cite{HarkoLobo2010}:
\begin{gather}
3\beta\big[f'R_{\mu\nu}+(g_{\mu\nu}\Box-\nabla_\mu\nabla_\nu)f'\big]-\tfrac12fg_{\mu\nu}\nonumber\\
=\tfrac{\alpha}{2}f'\big(T_{\mu\nu}-g_{\mu\nu}\mathcal L_m\big),
\label{eq:fRLm}
\end{gather}
with $f=f^{(r)}(y)$, $f'=f^{(r)\prime}(y)$ and $T_{\mu\nu}=-\frac{2}{\sg}\frac{\delta(\sg\mathcal L_m)}{\delta g^{\mu\nu}}$. Taking the divergence and using the Bianchi identity gives
\begin{equation}
\nabla^\mu T_{\mu\nu}=\big(g_{\mu\nu}\mathcal L_m-T_{\mu\nu}\big)\nabla^\mu\ln f'(y).
\label{eq:nonconservation}
\end{equation}
For Tsallis ($r=0$) one has $f'=1$, and Eq.~\eqref{eq:fRLm} is Einstein's equation with a conserved $T_{\mu\nu}$. R\'enyi, with $f'=(1-y/\cN)^{-1}$, and Sharma--Mittal, with $f'=(1-y/\cN)^{-r}$, instead predict an extra force on matter~\cite{Bertolami2007} through Eq.~\eqref{eq:nonconservation}, which in the weak-field regime is suppressed by $\nabla\ln f'\simeq(r/\cN)\nabla y$ with $y\sim\rho/\rho_P$.

\section{Cosmological constant from a volume constraint}\label{sec:Lambda}
With $\Phi=0$, the GfE action vanishes on Minkowski space and contains no bare cosmological constant. Its emergent term $\LG$ is not a constant cosmological term on shell (Sec.~\ref{sec:equiv}). \rev{In $\Lambda$-gravity from entropy~\cite{BianconiLambda2026}, a Lagrange multiplier $\Lambda$ defines the local volume of the theory; it adds a constant contribution to the dynamical $\LG$ and can be identified with the cosmological constant at low energy and small curvature.} Here we extend that construction to the three entropy families.

\subsection{Volume constraint for every entropy}
\rev{We implement the volume constraint in the spirit of Ref.~\cite{BianconiLambda2026}:} introduce an auxiliary scalar density $\varpi$ of weight one and hold it fixed when varying the metric. The constraint term is
\begin{gather}
\mathcal S_\Lambda=\frac1{\lP^4}\int\sg\,\mathcal L_\Lambda\,\dr\quad(d=4),\nonumber\\
\mathcal L_\Lambda=\frac{6\beta\Lambda}{\varpi}\big(\varpi-\sg\big),
\label{eq:LLambda}
\end{gather}
Here $\Lambda$ enforces $\varpi=\sg$; we use $\varpi$ for the volume density to distinguish it from the 1-form $\omega_\mu$. Since this term is independent of the entropy choice, it can be added to every member of Eq.~\eqref{eq:family}. It vanishes on the constraint surface, but its metric variation at fixed $\varpi$ does not:
\begin{equation}
\delta\big(\sg\,\mathcal L_\Lambda\big)\big|_{\varpi=\sg}=3\beta\Lambda\,\sg\,g_{\mu\nu}\,\delta g^{\mu\nu}.
\label{eq:dLLambda}
\end{equation}
Indeed, at fixed $\varpi$, $\partial[\sg(1-\sg/\varpi)]/\partial\sg=-1$ when $\varpi=\sg$, and $\delta\sg=-\frac12\sg\,g_{\mu\nu}\delta g^{\mu\nu}$. Because Eq.~\eqref{eq:modEin} is the metric equation divided by $\beta$, the constrained equation is
\begin{equation}
R^{\mathcal G}_{(\mu\nu)}-\tfrac12g_{\mu\nu}\Big(\RG-\frac V\beta\Big)+3\Lambda g_{\mu\nu}+\mathcal D_{(\mu\nu)}=\kappa_m\mathcal T_{(\mu\nu)}.
\label{eq:modEinL}
\end{equation}
For $\tGc=\tI$, Eq.~\eqref{eq:modEin} reduces to $3G_{\mu\nu}=3\cdot8\pi G\,T_{\mu\nu}$. Accordingly, Eq.~\eqref{eq:modEinL} becomes $G_{\mu\nu}+\Lambda g_{\mu\nu}=8\pi G\,T_{\mu\nu}$ in this limit, for all three entropy choices. Although $\varpi$ is not varied, the divergence of Eq.~\eqref{eq:modEinL}, together with the generalized Bianchi identity and the matter and G-field equations, enforces $\partial_\mu\Lambda=0$, as in unimodular gravity~\cite{HenneauxTeitelboim1989,Unruh1989}.

\paragraph{Two vacuum terms.} With $V=2\beta\Lambda_{q,r}$ (Sec.~\ref{sec:potentials}), the terms proportional to $g_{\mu\nu}$ in Eq.~\eqref{eq:modEinL} combine to $3(\Lambda+\frac13\Lambda_{q,r})g_{\mu\nu}$. For GfE, $\Lambda_{q,r}=\LG$, so its contribution to the combined coefficient is $\LG/3$ \rev{(Sec.~\ref{sec:GfE-Gfield})}; for the deformations, it is the power-mean potential~\eqref{eq:Lqr}. This potential has the sign of $q$ for Tsallis and is non-negative for R\'enyi when $q>0$ and Sharma--Mittal when $q,r>0$. As in GfE, it contributes on shell only through curvature-squared terms. Thus, in the weak-field regime, the constant $\Lambda$ is the only cosmological constant.

\subsection{Trace-free and trace equations}
Every term proportional to $g_{\mu\nu}$, including $\Lambda$ and $V$, cancels from the trace-free field equation:
\begin{equation}
\big[R^{\mathcal G}_{(\mu\nu)}+\mathcal D_{(\mu\nu)}\big]^{\rm TF}=\kappa_m\,\mathcal T^{\rm TF}_{(\mu\nu)}.
\label{eq:TF}
\end{equation}
Thus the cancellation mechanism of Ref.~\cite{BianconiLambda2026}—vacuum stress proportional to $g_{\mu\nu}$ drops out of the trace-free equations—holds for all three deformations. Their entropy dependence enters through the G-field relation~\eqref{eq:escort}.

The field equations determine $\Lambda$ only through their trace. \rev{Reference~\cite{BianconiLambda2026} derives an equation for $\Lambda$ by enforcing local Weyl stationarity of the action modulo the volume constraint, and notes that the theory does not predict its exact value;} we do not repeat that derivation. For the present family, the trace of Eq.~\eqref{eq:modEinL} in $d=4$, with $\xi=0$, is
\begin{equation}
12\Lambda=\RG-\frac{2V}{\beta}-g^{\mu\nu}\mathcal D_{\mu\nu}+\kappa_m\,g^{\mu\nu}\mathcal T_{\mu\nu}.
\label{eq:Ltrace}
\end{equation}
Unlike Eq.~\eqref{eq:TF}, this equation depends on $F$ and $\Psi$ through $\RG$, $V$, and $\mathcal D$. To first order in $\beta$, in vacuum or when curvature--matter couplings of order $\alpha\beta$ are neglected, it follows from the action $3\beta R$ plus the quadratic terms~\eqref{eq:quad} and the constraint term. In four dimensions, the Weyl-squared term is traceless and the Gauss--Bonnet term is topological, leaving only the $R^2$ contribution:
\begin{equation}
\Lambda=\frac14\Big(R-\frac{\Box R}{m_0^2}+8\pi G\,T\Big)+O(\beta^2),
\label{eq:Lambda1}
\end{equation}
with $m_0$ the scalar mass in Eq.~\eqref{eq:masses}. At this order, $\RG-2V/\beta=3R+O(\beta^2)$, so the entropy dependence of the trace relation enters only through $m_0$, not through the spin-2 sector. In the GR limit $m_0\to\infty$, Eq.~\eqref{eq:Lambda1} is the trace of $G_{\mu\nu}+\Lambda g_{\mu\nu}=8\pi GT_{\mu\nu}$. On the ghost-free $q=0$ branch, the trace follows from Eq.~\eqref{eq:fRLm} with $f$ replaced by $f-6\beta\Lambda$.

\section{Discussion and conclusions}\label{sec:disc}
\subsection{Universality}\label{sec:univ}
The perturbative results of this paper and Ref.~\cite{PaperII} are organized by the Taylor coefficients of $F$ and $\Psi$ at the identity (Table~\ref{tab:universal}). The normalizations $F'(1)=\Psi'(0)=1$ fix the common Einstein limit. Through quadratic order, only $\kappa_2=-\frac12F''(1)$ and $\psi_2=\Psi''(0)$ enter [Eq.~\eqref{eq:quadgen}]. Cubic order introduces $\kappa_3:=\frac16F'''(1)$ and $\psi_3:=\Psi'''(0)$. In vacuum, $\TrF Y=-3\beta R$, so the $\psi_3(\TrF Y)^3/6$ contribution is proportional to $R^3$ and vanishes on Ricci-flat backgrounds. Accordingly, three coefficients suffice near such backgrounds and four are needed in general.

At fixed $q$, the families share quantities controlled by $F$, including the spin-2 mass, the Jacobson--Myers entropy constant, and leading black-hole corrections. Their differences enter through $\psi_2$ and $\psi_3$, which control inter-sector couplings. Nonperturbative predictions, including Planck-scale de Sitter vacua and inflation, depend on the full functions $F$ and $\Psi$. The domain and associated black-hole curvature bound are common to the family.

\begin{table*}[t]
\caption{Universal and entropy-dependent predictions for the family~\eqref{eq:family} with GfE curvature blocks. The controlling Taylor coefficients are shown; $\cN=11$ and $\beta'=1/48\pi$. The coefficients $c_I$ and $c_{RK}$ multiply $I_1:=R_{\mu\nu}{}^{\rho\sigma}R_{\rho\sigma}{}^{\eta\theta}R_{\eta\theta}{}^{\mu\nu}$ and $R\mathscr K$ in the black-hole effective action of Ref.~\cite{PaperII}; $\mathfrak s_0$ is the constant term in de Sitter horizon entropy. A dagger marks results from Ref.~\cite{PaperII}. GfE is $q=r=1$.}
\label{tab:universal}
\begin{ruledtabular}
\begin{tabular}{llccc}
Quantity & Controlled by & Tsallis & R\'enyi & Sharma--Mittal\\
\hline
Einstein limit: Newton's constant and leading matter couplings & $F'(1)=\Psi'(0)=1$ & \multicolumn{3}{c}{universal}\\
Domain of the action; curvature bound $M>M_P/\sqrt{96\pi}$$^\dagger$ & spectrum of $\tN$ & \multicolumn{3}{c}{universal}\\
Spin-2 mass $m_2^2$ & $\kappa_2$ & \multicolumn{3}{c}{$-6/(5q\beta)$}\\
Weight $\kappa_2$ of $\TrF\tX^2$; Jacobson--Myers entropy $4\pi\kappa_2\beta'/3$ & $\kappa_2$ & \multicolumn{3}{c}{$q/2$;\quad $q/72$}\\
Scalar mass $m_0^2$; $\Lambda$ trace sector at $O(\beta)$ & $\kappa_2,\psi_2$ & $\frac{3}{5q\beta}$ & $\frac{33}{\beta(27+28q)}$ & $\frac{33}{\beta(27r+28q)}$\\
Coupling of form sectors and of $R$ to $\mathcal L_m$ & $\psi_2$ & 0 & $\frac{1-q}{\cN}$ & $\frac{r-q}{\cN}$\\
Cubic black-hole coupling $c_I=\kappa_3/3$; $T_H$, Wald entropy$^\dagger$ & $\kappa_3$ & \multicolumn{3}{c}{$q(q+1)/18$}\\
$R\mathscr K$ coupling $c_{RK}=\psi_2\kappa_2$; horizon radius, ISCO$^\dagger$ & $\kappa_2,\psi_2$ & 0 & $\frac{q(1-q)}{2\cN}$ & $\frac{q(r-q)}{2\cN}$\\
de Sitter entropy constant $\mathfrak s_0=\frac1{48}\big(\frac{34}3\kappa_2+36\psi_2\big)$$^\dagger$ & $\kappa_2,\psi_2$ & $\frac{17q}{144}$ & $\frac{79q+108}{1584}$ & $\frac{79q+108r}{1584}$\\
$R^3$ term $\frac{\psi_3}{6}(\TrF Y)^3$ & $\psi_3$ & 0 & $\frac{2(1-q)^2}{\cN^2}$ & $\frac{(q-r)(2q-r-1)}{\cN^2}$\\
de Sitter vacua, inflation, strong fields$^\dagger$ & full $F$, $\Psi$ & \multicolumn{3}{c}{member dependent}\\
\end{tabular}
\end{ruledtabular}
\end{table*}

\subsection{Summary}
We have extended GfE to Tsallis, R\'enyi, and Sharma--Mittal relative entropies\rev{, which define gravity from generalized entropy (GfGE)}. Their common normalization fixes the Einstein limit, while a finite set of Taylor coefficients controls the leading higher-curvature terms, extra modes, G-field potential, and cosmological trace equation. The admissible index range is $q\ge0$ with real $r$; although the associated divergences satisfy the stated non-negativity conditions, the action itself is sign-indefinite because it includes a linear term.

In GfE, the linearized spectrum contains a healthy scalar and a Planck-scale tachyonic spin-2 ghost\rev{, as first found by Pereira~\cite{Pereira2026a}; such ghosts also occur} in quadratic gravity~\cite{Stelle1977,Stelle1978,LuPope2011}. The massive spin-2 mode is absent only in the trace-projected or flat-$F$ branches. At $q=0$, Tsallis is exactly general relativity; R\'enyi and Sharma--Mittal instead give ghost-free $f(R,\mathcal L_m)$ theories with a scalar G-field and nonconserved matter stress tensor. On shell, the G-field is proportional to the Wald tensor, so horizon entropy is weighted by its normal components. Analyticity then gives area, constant, and inverse-area terms for single-scale horizons, rather than the postulated power laws. The volume-constraint construction extends to all three families, with the scalar mass controlling the leading entropy dependence of the trace equation.

Ref.~\cite{PaperII} studies black holes, cosmological solutions, and observational consequences; a separate analysis will compare the information-theoretic meaning of the three entropies.

\subsection{Open problems}

\paragraph{Effective field theory and ghosts.} If GfE is an effective field theory below $|m_2|\simeq13\,M_P$~\cite{Donoghue1994}, quantum corrections generate higher-curvature terms, including the two-loop contribution cubic in the Riemann tensor~\cite{GoroffSagnotti1986}. In vacuum, metric field redefinitions can remove the curvature-squared terms; with matter, they become contact interactions, making the curvature--matter couplings of Sec.~\ref{sec:spectrum} basis dependent. If instead the action is fundamental, a nonlinear Hamiltonian analysis is needed to determine the fate of the spin-2 ghost.

\paragraph{Dark matter.} Reference~\cite{GfE} suggests that the G-field may contribute to dark matter. Around flat space, its fluctuations have the masses in Eq.~\eqref{eq:masses}; they are Planckian in GfE, and much lighter modes require $q\gg1$. But the curvature-expansion coefficients also grow with $q$ (for example, $\kappa_2=q/2$), so the effective expansion parameter becomes $q\beta\mathcal R$. The weak-coupling analysis must therefore be revisited in this regime.

\paragraph{Cosmological constant.} The equation for $\Lambda$ \rev{obtained from local} Weyl stationarity modulo the volume constraint~\cite{BianconiLambda2026} remains to be worked out for general $F$ and $\Psi$ beyond first order in $\beta$ [Eq.~\eqref{eq:Ltrace}]. The interplay between $\Lambda$ and the matter couplings on the $q=0$ branch also requires study.

\paragraph{Lorentzian positivity.} The Lorentzian positivity conditions for non-integer powers of $\tN$ (complex or Jordan spectra) deserve a systematic study.

\begin{acknowledgments}
\rev{The author thanks Ginestra Bianconi for reading the manuscript and for helpful comments on the interpretation of gravity from entropy. Claude (Anthropic) was used as a tool for drafting and editing the text and for symbolic and numerical checks. The author reviewed and verified all derivations and results and takes full responsibility for the content of this paper.}
\end{acknowledgments}

\appendix
\section{Weyl decomposition and trace identities}\label{app:weyl}
In $d=4$, with the Kulkarni--Nomizu product
\begin{equation}
(A\owedge B)_{\mu\nu\rho\sigma}=A_{\mu\rho}B_{\nu\sigma}+A_{\nu\sigma}B_{\mu\rho}-A_{\mu\sigma}B_{\nu\rho}-A_{\nu\rho}B_{\mu\sigma},
\end{equation}
the Riemann tensor decomposes as
\begin{equation}
R_{\mu\nu\rho\sigma}=C_{\mu\nu\rho\sigma}+\tfrac12(S\owedge g)_{\mu\nu\rho\sigma}+\tfrac{R}{24}(g\owedge g)_{\mu\nu\rho\sigma},
\end{equation}
with $C$ completely traceless. Raising the last index pair defines the bivector operators $\mathsf C$, $\mathsf S$, and $\mathsf R_0$. Since $\mathsf R_0=\frac R6\mathbf 1$ and $(g\owedge g)_{\mu\nu}{}^{\rho\sigma}=4\mathbf 1_{\mu\nu}{}^{\rho\sigma}$, their traces are full index contractions, as in Sec.~\ref{sec:topo}. Tracelessness of $C$ and $S$ gives
\begin{gather}
\TrM\mathsf C=\TrM\mathsf S=0,\qquad \TrM\mathsf R_0=R,\\
\TrM\mathsf C\mathsf S=\TrM\mathsf C\mathsf R_0=\TrM\mathsf S\mathsf R_0=0,\\
\TrM\mathsf C^2=C^2,\qquad \TrM\mathsf S^2=2S_{\mu\nu}S^{\mu\nu},\\ \TrM\mathsf R_0^2=\tfrac16R^2 .
\end{gather}
Summing these traces yields $R_{\mu\nu\rho\sigma}R^{\mu\nu\rho\sigma}=C^2+2S_{\mu\nu}S^{\mu\nu}+\frac16R^2$ and $R_{\mu\nu}R^{\mu\nu}=S_{\mu\nu}S^{\mu\nu}+\frac14R^2$. Hence
\begin{gather}
\mathcal E=C^2-2S_{\mu\nu}S^{\mu\nu}+\tfrac16R^2,\\
R_{\mu\nu}R^{\mu\nu}=\tfrac12C^2+\tfrac13R^2-\tfrac12\mathcal E .\label{eq:ric2}
\end{gather}
For the blocks~\eqref{eq:blocks},
\begin{gather*}
\TrM\mathbf X_{(1)}^2=\gamma_{S1}^2S_{\mu\nu}S^{\mu\nu}+\tfrac14\gamma_{R1}^2R^2,\\
\TrM\mathbf X_{(2)}^2=\gamma_W^2C^2+2\gamma_{S2}^2S_{\mu\nu}S^{\mu\nu}+\tfrac16\gamma_{R2}^2R^2,
\end{gather*}
and eliminating $S_{\mu\nu}S^{\mu\nu}=\frac12(C^2-\mathcal E)+\frac1{12}R^2$ gives Eqs.~\eqref{eq:quad} and~\eqref{eq:Q}.

\paragraph*{Curvature traces of the GfE blocks.} For the GfE blocks of Sec.~\ref{sec:topo}, the identities $\TrF\tX=3R$ and~\eqref{eq:TrX2} hold in any dimension. The scalar and vector blocks give $\TrM\mathbf X_{(0)}=R$ and $\TrM\mathbf X_{(1)}=R_{\mu\nu}g^{\nu\mu}=R$. In the 2-form block, Eq.~\eqref{eq:g2} and antisymmetry imply
\begin{equation}
R_{\mu\nu\rho\sigma}[g_{(2)}]^{\rho\sigma\mu\nu}=\tfrac12\big(R_{\mu\nu}{}^{\mu\nu}-R_{\mu\nu}{}^{\nu\mu}\big)=R .
\end{equation}
The kernel of $\mathbf X_{(2)}$ is $R_{\mu\nu\rho\sigma}[g_{(2)}]^{\rho\sigma\eta\theta}=R_{\mu\nu}{}^{\eta\theta}$. Pair symmetry then gives $\TrM\mathbf X_{(2)}^2=R_{\mu\nu}{}^{\eta\theta}R_{\eta\theta}{}^{\mu\nu}=R_{\mu\nu\rho\sigma}R^{\mu\nu\rho\sigma}$. The scalar and vector blocks contribute $R^2$ and $R_{\mu}{}^\nu R_{\nu}{}^\mu$, respectively.

\section{Linearized spectrum}\label{app:lin}
Consider $\mathcal L=R+aR_{\mu\nu}R^{\mu\nu}+bR^2$ expanded about Minkowski space. Using Eq.~\eqref{eq:ric2}, the coefficients of $C^2$ and $R^2$ are $a_C=a/2$ and $a_R=b+a/3$.

\paragraph*{Spin 2.} Let $g=\eta+h$ in transverse-traceless gauge, $\partial^\mu h_{\mu\nu}=0=h^\mu{}_{\mu}$. The linearized Ricci tensor
\begin{equation}
R^{(1)}_{\mu\nu}=\tfrac12\big(\partial_\rho\partial_\mu h^\rho{}_\nu+\partial_\rho\partial_\nu h^\rho{}_\mu-\Box h_{\mu\nu}-\partial_\mu\partial_\nu h\big)
\end{equation}
reduces to $-\frac12\Box h_{\mu\nu}$, while $R^{(1)}=0$. In this gauge the second-order Einstein--Hilbert density, the Fierz--Pauli Lagrangian~\cite{FierzPauli1939}, is $(\sg R)^{(2)}=-\frac14\partial_\lambda h_{\mu\nu}\partial^\lambda h^{\mu\nu}\simeq\frac14h^{\mu\nu}\Box h_{\mu\nu}$. The curvature-squared terms begin at second order, so only the linearized curvatures contribute, and
\begin{equation}
\mathcal L^{(2)}_{TT}\simeq\tfrac14h^{\mu\nu}\Box(1+a\Box)h_{\mu\nu}.
\end{equation}
The massive pole therefore has $m_2^2=-1/a=-1/(2a_C)$.

\paragraph*{Spin 0.} Let $g=e^{2\vartheta}\eta$. In $d=4$,
\begin{gather}
R_{\mu\nu}=-2\partial_\mu\partial_\nu\vartheta-\eta_{\mu\nu}\Box\vartheta+O(\vartheta^2),\\
\sg R=-6e^{2\vartheta}\big(\Box\vartheta+(\partial\vartheta)^2\big).
\end{gather}
Integrating by parts gives $(\sg R)^{(2)}\simeq6(\partial\vartheta)^2\simeq-6\vartheta\Box\vartheta$, and
\begin{gather}
(R_{\mu\nu}R^{\mu\nu})^{(2)}=4(\partial_\mu\partial_\nu\vartheta)^2+8(\Box\vartheta)^2\simeq12(\Box\vartheta)^2,\\
(R^2)^{(2)}=36(\Box\vartheta)^2 ,
\end{gather}
so that
\begin{equation}
\mathcal L^{(2)}\simeq-6\vartheta\Box\big[1-(2a+6b)\Box\big]\vartheta,
\end{equation}
i.e.\ $m_0^2=1/(2a+6b)=1/(6a_R)$. The massless conformal factor is nonpropagating and has the usual wrong-sign kinetic term. The massive pole has the opposite residue and is healthy when $m_0^2>0$.

We verified both results by expanding the exact curvature of $g=\mathrm{diag}(-1,1+\epsilon f,1-\epsilon f,1)$ and of $g=e^{2\epsilon\vartheta}\eta$ to $O(\epsilon^2)$ with computer algebra; the resulting Euler--Lagrange equations are $\Box f+a\Box^2f=0$ and $\Box\vartheta-(2a+6b)\Box^2\vartheta=0$. Combining these results with Eq.~\eqref{eq:quad} gives Eq.~\eqref{eq:masses}.

\paragraph*{Sharma--Mittal masses.} For the GfE blocks, $\beff=3\beta$, $\mathcal W=\frac52\beta^2$, and $\mathcal Q=\frac53\beta^2$. Substituting $\kappa_2=q/2$ and $\psi_2=(r-q)/\cN$ into Eq.~\eqref{eq:masses} gives, keeping $\cN$ symbolic,
\begin{equation}
m_0^2=\Big\{2\beta\Big[\frac{5q}{6}+\frac{9(r-q)}{2\cN}\Big]\Big\}^{-1}.
\label{eq:m0gen}
\end{equation}
For $\cN=11$, this reduces to Eq.~\eqref{eq:massesSM}; the spin-2 mass, $m_2^2=-\beff/(2\kappa_2\mathcal W)=-6/(5q\beta)$, is independent of $r$.

\section{Metric variation of the dressed action}\label{app:var}
Since the auxiliary fields $\tTh$ and $\tGc$ enter $\tilde{\mathcal L}$ algebraically, the on-shell metric equations of the G-field action coincide with those of the original Lagrangian $\mathcal L(g,R_{\mu\nu\rho\sigma},\Phi)$. The choice of the components of $\tGc$ held fixed in the metric variation [assumption~(\ref{as:Gfield})] changes the off-shell form of the equations only by terms proportional to the auxiliary equations~\eqref{eq:aux}. For a Lagrangian that depends on the metric and the Riemann tensor, the gravitational part of the field equations is~\cite{Padmanabhan2013}
\begin{gather}
E_{\mu\nu}=P_{(\mu}{}^{\lambda\eta\theta}R_{\nu)\lambda\eta\theta}-\tfrac12g_{\mu\nu}\mathcal L-2\nabla^\lambda\nabla^\eta P_{(\mu|\lambda\eta|\nu)},\nonumber\\
P^{\mu\nu\rho\sigma}:=\frac{\partial\mathcal L}{\partial R_{\mu\nu\rho\sigma}}.
\label{eq:Padm}
\end{gather}
The curvature enters $\mathcal L$ only through $\tN=\tI+\alpha\tM\tg^{-1}-\beta\tX$; in vacuum, or for $\xi=0$, only through $-\beta\tX$. In that case, by the chain rule and Eq.~\eqref{eq:Ggen},
\begin{equation}
\frac{\partial\mathcal L}{\partial\tX}=\beta\,\Omega F'(\tN)=\beta\,\tGc\quad\text{(on shell)},
\end{equation}
block by block. Using $R=g^{\mu\rho}g^{\nu\sigma}R_{\mu\nu\rho\sigma}$, $R_\nu{}^\mu=g^{\lambda\rho}g^{\mu\sigma}R_{\lambda\nu\rho\sigma}$, and $\mathbf X_{(2)}=R_{\mu\nu}{}^{\rho\sigma}$, we find
\begin{equation}
P^{\mu\nu\rho\sigma}=\beta\,\mathbb P\Big[\bm{\mathcal G}_{(0)}\,g^{\mu\rho}g^{\nu\sigma}+g^{\mu\rho}\bm{\mathcal G}_{(1)}{}^{\nu\sigma}+\bm{\mathcal G}_{(2)}{}^{\mu\nu\rho\sigma}\Big],
\label{eq:PG}
\end{equation}
with $\mathbb P$ as in Sec.~\ref{sec:equiv} and all indices raised with $g$. The G-field is thus, up to the factor $\beta$, the tensor $\partial\mathcal L/\partial R_{\mu\nu\rho\sigma}$ that enters both the field equations and the Wald entropy. Dividing Eq.~\eqref{eq:Padm} by $\beta$ and using $\mathcal L=\beta\RG-V-\alpha\MG$ on shell gives Eq.~\eqref{eq:modEin}, with
\begin{equation}
R^{\mathcal G}_{\mu\nu}:=\frac{P_{\mu}{}^{\lambda\eta\theta}R_{\nu\lambda\eta\theta}}{\beta},\quad \mathcal D_{\mu\nu}:=-\frac{2\nabla^\lambda\nabla^\eta P_{\mu\lambda\eta\nu}}{\beta}.
\label{eq:RGD}
\end{equation}
Reference~\cite{GfE} writes the same equations in terms of a dressed Ricci tensor and explicit derivative terms of the G-field. Both forms are Euler--Lagrange equations of classically equivalent actions (Sec.~\ref{sec:equiv}) and therefore agree on shell.
\paragraph*{Checks.} For the scalar block, $P^{\mu\nu\rho\sigma}=\frac12\beta\bm{\mathcal G}_{(0)}(g^{\mu\rho}g^{\nu\sigma}-g^{\mu\sigma}g^{\nu\rho})$. It yields $R^{\mathcal G}_{\mu\nu}=\bm{\mathcal G}_{(0)}R_{\mu\nu}$ and $\mathcal D_{\mu\nu}=(g_{\mu\nu}\Box-\nabla_\mu\nabla_\nu)\bm{\mathcal G}_{(0)}$, the standard $f(R)$ equations~\cite{SotiriouFaraoni2010} when $\bm{\mathcal G}_{(0)}=\partial_Rf/\beta$. At $\tGc=\tI$, each block contributes $\frac12\beta(g^{\mu\rho}g^{\nu\sigma}-g^{\mu\sigma}g^{\nu\rho})$, giving the value of $P^{\mu\nu\rho\sigma}$ for $3\beta R$, with $R^{\mathcal G}_{\mu\nu}=3R_{\mu\nu}$ and $\mathcal D_{\mu\nu}=0$, consistent with $\RG=3R$. Finally, $\MG$ supplies $\kappa_m\mathcal T_{\mu\nu}$, while $V$ contributes only through the term $-\frac12g_{\mu\nu}\mathcal L$, recovering Eq.~\eqref{eq:modEin}.

\section{Derivation of the power-mean cosmological term}\label{app:lambda}
For $q,r\neq0$, define $\varepsilon=1-1/q$ as in Eq.~\eqref{eq:Lqr}. Let $\bar Z:=\av{\tTh^{1-q}}$ denote $\bar Z_q$ evaluated at the stationary point. Equation~\eqref{eq:escort} gives $\tTh=(\bar Z^{\,1-s}\tGc)^{-1/q}$, and hence
\begin{equation}
\tTh^{1-q}=(\bar Z^{\,1-s}\tGc)^{\varepsilon}\ \Rightarrow\ \bar Z=\bar Z^{(1-s)\varepsilon}\av{\tGc^\varepsilon}.
\end{equation}
Since $1-(1-s)\varepsilon=r/q$, we obtain $\bar Z=\av{\tGc^\varepsilon}^{q/r}$. The remaining terms are
\begin{gather}
\TrF\tGc\tTh=\bar Z^{\,s-1}\TrF\tTh^{1-q}=\cN\bar Z^{\,s},\\
L(\tTh)=\frac{\cN}{r-1}(\bar Z^{\,s}-1).
\end{gather}
Inserting these into $V=\TrF\tGc-\TrF\tGc\tTh-L(\tTh)$ gives
\begin{equation}
V=\TrF\tGc-\cN-\frac{r\cN}{r-1}\big(\bar Z^{\,s}-1\big).
\end{equation}
Finally, $\bar Z^{\,s}=\av{\tGc^\varepsilon}^{qs/r}=\mathsf M_\varepsilon^{\varepsilon qs/r}=\mathsf M_\varepsilon^{(r-1)/r}$, while $\frac r{r-1}(x^{(r-1)/r}-1)=\ln_{1/r}x$, which yields Eq.~\eqref{eq:Lqr}. The limits $r\to1$ and $q\to1$ follow from $\ln_1=\ln$ and $\mathsf M_\varepsilon\to\exp\av{\ln\tGc}$ as $\varepsilon\to0$.

\section{Symbolic and numerical checks}\label{app:num}
We used SymPy, NumPy, and SciPy for the following checks. The rank-one identity~\eqref{eq:rank1}, curvature traces~\eqref{eq:TrX2}, Weyl decomposition, and quadratic form~\eqref{eq:quad} were checked for random Lorentzian metrics, algebraic curvature tensors, and couplings. Exact curvature expansions were used to verify the linearized equations in Appendix~\ref{app:lin}. We also checked the expansion~\eqref{eq:SMexp}, derivatives~\eqref{eq:Psider}, and identities~\eqref{eq:frenyi} and~\eqref{eq:skew}. For random spectra and $(q,r)\in\{(0.5,0.3),(1.7,2.5),(0.8,1),(1.3,1.3),(2,0.4)\}$, we checked Eq.~\eqref{eq:Lqr} and the Bregman identity. Finally, we compared the volume-constrained equations~\eqref{eq:modEinL} and~\eqref{eq:Lambda1} with those obtained from $R+R^2/(6m_0^2)$ gravity.

\bibliography{GfE_paperI}
\end{document}